\documentclass[11pt,onecolumn,amssymb,nofootinbib,floatfix]{revtex4}
\usepackage{amsmath, amsthm, amscd, amssymb}
\usepackage{graphicx, braket}
\usepackage{bm}
\usepackage{bbm}
\usepackage{epstopdf}
\usepackage{capt-of}
\begin{document}

\title{\bf Modeling the flyby anomalies with dark matter scattering: update with additional data and further predictions}
\author{Stephen L. Adler}
\email{adler@ias.edu} \affiliation{Institute for Advanced Study,
Einstein Drive, Princeton, NJ 08540, USA.}

\begin{abstract}
We continue our exploration of whether the flyby anomalies can be explained by
scattering of spacecraft nucleons from dark matter gravitationally bound to the earth,
with the addition of data from five new flybys to that from the original six.
We continue to use our model in which inelastic and elastic scatterers
populate shells generated by the precession of circular orbits with normals tilted
with respect to the earth's axis. With 11 data points and 8 parameters in the model,
a statistically meaningful fit is obtained with a chi-squared of 2.7.  We give plots
of the anomalous acceleration along the spacecraft trajectory, and the cumulative
velocity change, for the five flybys which exhibit a significant nonzero anomaly.  We also discuss
implications of the fit for dark matter-nucleon cross sections, give the prediction of our fit
for the anomaly to be expected from the future Juno flyby, and give predictions of our fit for flyby
orbit orientation changes. In addition we give formulas for estimating the flyby temperature increase
caused by dark matter inelastic scattering, and for the fraction of flyby nucleons undergoing such scatters.
Finally, for circular satellite orbits, we give a table of predicted secular
changes in orbit radius.  These are much too large to be reasonable -- comparing with data for COBE and
GP-B supplied to us by Edward Wright (after the first version of this paper was posted), we find that our
model predicts changes in orbit radius that are too large by many orders of magnitude. So the model
studied here is ruled out.  We conclude that further modeling of the flyby anomalies must simultaneously
attempt to fit constraints coming from satellite orbits.

\end{abstract}

\maketitle

\section{Introduction}
In this paper we follow up our earlier investigations \cite{adler1}, \cite{adler2} of the anomalous geocentric frame
orbital energy changes that are observed during earth flybys of various spacecraft, as
reported by Anderson et al. \cite{anderson}.   Some flybys show energy decreases, and
others energy increases, with the largest anomalous velocity changes of order 1 part
in $10^6$.  While the possibility that these anomalies are artifacts of the orbital
fitting method used in \cite{anderson} is still being actively explored, there is also
a chance that they may represent new physics.

In \cite{adler1} we explored the possibility that the flyby anomalies
result from scattering of spacecraft nucleons from dark matter particles in orbit around the earth, with the
observed velocity decreases arising from elastic scattering, and the observed velocity
increases arising from exothermic inelastic scattering, which can impart an energy impulse to a spacecraft nucleon.  Many
constraints on this hypothesis were analyzed in \cite{adler1}, with the conclusion that the dark matter
scenario is not currently ruled out, but requires dark matter to be non-self-annihilating, with the dark
matter scattering cross section on nucleons much larger, and the dark matter mass much lighter, than
usually assumed.  These constraints on the dark matter scattering cross section and mass will be revisited,
and somewhat weakened, in the present paper.

In \cite{adler2} we constructed a model for the spatial and velocity distribution functions for dark matter
particles in earth orbit, based on assuming
two populations of dark matter particles, one of which scatters on nucleons elastically, and the other of which scatters inelastically, with each population
taken to fill a shell-like distribution of  orbits generated by the precession of a
tilted circular orbit around the earth's rotation axis.  Fit of this model to the data were presented, but since the model has 8 parameters, and
there were only 6 data points in the original flyby set, the statistical significance of the fits was doubtful.  After \cite{adler2} was written,
Anderson and Campbell \cite{andersoncampbell}  furnished us with data for 5 additional flybys, all of which show no velocity anomaly.  In the
present paper, we add these 5 flybys to the original 6 and  fit the model of \cite{adler2} to all of the data points.  We find that with
11 data points and 8 parameters, we get a statistically meaningful fit with a chi-squared of 2.7.  In this fit, the dark matter shells lie much
closer to earth than they did in the fits exhibited in \cite{adler2}.

The formulas defining the model were given in Sec. II of \cite{adler2} (with further detail in the Appendices of the arXiv version), and will not be repeated here.
So reading the earlier paper is a necessary prerequisite for fully understanding the model analyzed in this paper.
Our focus here is on presenting the orbital parameters of the new flybys and the results of the fit to the full 11 flyby set.  In addition to giving the
fits to the total velocity anomaly, we give the contributions of the shells of elastic and inelastic scatterers to this total. We also plot the
anomalous acceleration along the spacecraft trajectory and the cumulative velocity change, for the five flybys which show significant nonzero velocity
anomalies,  give constraints on dark matter-nucleon cross sections implied by our fit, and give formulas for estimating the flyby temperature
change and the fraction of nucleons that undergo exothermic dark matter scatters.  For the future Juno flyby, our 11 flyby fit predicts an anomaly of 11.6 mm/s.

We conclude by giving predictions of our model for flyby orbit orientation anomalies, and for secular changes in orbit radius of satellites in circular orbits.
The latter, obtained from a program that can treat non-circular orbits as well, shows that the predicted orbit radius changes for COBE and GP-B are orders
of magnitude larger than observational data supplied to us by Edward Wright after the first version of this paper was posted to arXiv: see the Added note to
Section VII.
Hence the model studied here is ruled out, and any further fitting of more general
forms of the dark matter model, or of other types of models, to the flyby anomalies, must simultaneously attempt to fit constraints coming from closed
satellite orbits.  This greatly complicates the analysis, and we shall pursue this further only if the future Juno flyby, which has orbital parameters
similar to GLL-I, confirms the presence of a kinetic energy anomaly.

 \section{Flyby orbital plane parameters}

 As in \cite{adler2},  it will be convenient for to carry out all flyby orbit calculations in
 the flyby orbital plane.  Let $x_o,y_o,z_o$ be a Cartesian axis system, with $z_o$ normal to the flyby orbital
 plane.   The flyby orbit can then be written in parametric form as
 \begin{align}\label{eq:param}
 x_o(t)=&r(t) \cos \theta_o(t)~,~~~y_o(t)=r(t) \sin \theta_o(t)~~~,\cr
 r(t)=&\frac {p} {1+e \cos \theta_o(t)},~~~~~~R_f=\frac{p}{1+e}~~~,\cr
 dx_o(t)/dt=& \frac {-V_f \sin \theta_o(t)}{1+e}=\frac{-y_o(t)}{1+e\cos \theta_o(t)} d\theta_o(t)/dt~~~,\cr
 dy_o(t)/dt=&\frac{V_f (e+\cos \theta_o(t))}{1+e}=\frac{e\,r(t)+x_o(t)}{1+e \cos \theta_o(t)} d\theta_o(t)/dt~~~,\cr
 d\theta_o(t)/dt=&\frac{R_fV_f}{r(t)^2}=\frac{(GM_{\oplus}p)^{1/2}}{r(t)^2}~~~.\cr
 \end{align}
The scale parameter $p$, the eccentricity $e$, the velocity at closest approach to earth $V_f$,
the radius at closest approach $R_f$, and the velocity at infinity $V_{\infty}$ are given in Table I  for the six original flybys, and
in Table II for the five new ones and the future Juno flyby,
together with the polar angle $I$ and azimuthal angle $\alpha$ of the earth's north pole with respect to the
$x_o,y_o,z_o$ coordinate system.  Formulas for obtaining $R_f$, $p$, $e$, and $\alpha$ from the data presented in \cite{anderson}
and \cite{andersoncampbell} are given in \cite{adler2}.

\begin{table} [t]\label{table:param1}
\centering
\caption{Orbital parameters for the original six flybys}
\begin{tabular} {|c| c |c| c| c| c | c|}
\hline\hline
~~~&~~~GLL-I~~~ & ~~~GLL-II~~~ & ~~~NEAR~~~ & ~~~Cassini~~~ & ~~~Rosetta~~~ & ~~~Messenger~~~ \\
\hline
$V_f$ (km/s) & 13.740 & 14.080 & 12.739 & 19.026 & 10.517 & 10.389\\
$R_f$ (km)   &7,334  & 6,674 & 6,911 & 7,544 & 8,332 & 8,715  \\
$V_{\infty}$ (km/s)& 8.949 & 8.877 & 6.851 & 16.010 & 3.863 & 4.056 \\
$e$ & 2.474 & 2.320 & 1.814 & 5.851 & 1.312 & 1.360\\
$p$ (km) & 25,480 & 22,160 & 19,450 & 51,690 & 19,260 &  20,570 \\
$I$ (deg) & 142.9 & 138.7 & 108.0 & 25.4 & 144.9 & 133.1 \\
$\alpha$ (deg) & -45.1 & -147.4 & -55.1 & -158.4 & -53.1 & 0.0 \\
\hline
\end{tabular}
\end{table}

\begin{table} [b]\label{table:param2}
\centering
\caption{Orbital parameters for the five new flybys, and the future Juno flyby}
\begin{tabular} {|c| c |c| c| c| c ||c|}
\hline\hline
~~~&~~~Rosetta II~~~ & ~~~Rosetta III~~~ & ~~~EPOXI I~~~ & ~~~EPOXI II~~~ & ~~~EPOXI III~~~& ~~~Juno~~~ \\
\hline
$V_f$ (km/s) & 12.49 & 13.34 & 6.94 & 5.29 & 5.73&15.024 \\
$R_f$ (km)   &11,656.6  & 8,860.62 & 21,985.02 & 49,786.17 & 36,775.84&6871.48  \\
$V_{\infty}$ (km/s)& 9.36 & 9.38 & 3.45 & 3.46 & 3.34 &10.474\\
$e$ & 3.562 & 2.956 & 1.656 & 2.495 & 2.029&2.891 \\
$p$ (km) & 53,177.89 & 35,051.16 & 58,402.95 & 174,016.7 & 111,402.94 &26738.31 \\
$I$ (deg) & 115.08 & 155.63 & 19.83 & 92.81 & 103.07 &48.23 \\
$\alpha$ (deg) & 175.35 & 110.04 &158.28 & 139.34 & -142.29&140.01 \\
\hline
\end{tabular}
\end{table}

\section{Model parameters and fit to 11 flybys}

The model of \cite{adler2} postulates two shells of dark matter gravitationally bound to earth.  An inelastic scatterer shell is the locus obtained by starting with a circular
orbit with normal  tilted at angle $\psi_i$ with respect to the earth's rotation axis, and smearing this orbit into a uniform shell by rotation around the earth's rotation axis
to represent the effect of quadrupole moment-induced precession.  The radial density in the shell is assumed to have a Gaussian profile proportional to $e^{-(r-R_i)^2/D_i^2}$,
and the effective shell density times scattering cross section on flyby nucleons is denoted by $\rho_i$.  This gives four parameters for the inelastic shell, the tilt angle
$\psi_i$, the central shell radius $R_i$, the shell width $D_i$, and the effective density $\rho_i$.  For the elastic scatterer shell (which could arise as daughter products from
interactions of the inelastic scatterers with nucleons) we assume an analogous structure, introducing four more parameters, again a tilt angle $\psi_e$, a central radius $R_e$,
a shell width $D_e$, and an effective density $\rho_e$.  A more detailed description of these parameters is given in Section 2.6 of \cite{adler2} (Section IIF of the arXiv version).

We note that the closest approach to earth of the flybys in our fit (flyby GLL-II) is $R_f=6,674$ km, corresponding to an altitude $H=303$ km, and thus the fits would be unaltered if the shell radial profile were changed to a cut-off Gaussian  $e^{-(r-R_i)^2/D_i^2}\theta(r-6,674 {\rm km})$, with $\theta(r)$ the usual step function.   Hence the model cannot be used to make predictions for orbits that lie below an altitude of 303 km, which includes the majority of low earth orbit satellites.  Predictions of the model for earth orbiting satellites in circular orbits lying above the cut-off will be given below.

 As in \cite{adler2}, fitting this model to the data was carried out by minimizing a least squares likelihood function $\chi^2$, defined as
 \begin{equation}\label{eq:chisq}
 \chi^2=\sum_{k=1}^{11} (\delta v_{k; {\rm th}}-\delta v_{k; {\rm A}})^2/\sigma_{k; \rm A}^2~~~,
 \end{equation}
 where $k$ indexes the six original flybys  discussed in \cite{anderson} and the five new flybys reported in \cite{andersoncampbell}, where the  $\delta v_{k; {\rm th}}$ are the
 theoretical values of the velocity anomalies computed from our model, the $\delta v_{k; {\rm A}}$ are
 the observed values for these anomalies reported in \cite{anderson} and \cite{andersoncampbell}, and the $\sigma_{k; \rm A}$ are the corresponding
 estimated errors in these anomalies.  Since the quoted $\sigma_{k; {\rm A}}$ values contain
 both systematic and statistical components, a least squares likelihood function is not a true statistical chi square
 function, but having a quadratic form is very convenient for the following reason. Because the theoretical values
 $\delta v_{k; {\rm th}}$ are linear in the dark matter density times cross section parameters $\rho_{i,e}$,
 \begin{equation}\label{eq:linear}
 \delta v_{k; {\rm th}} = \rho_i \delta v_{k; i} + \rho_e \delta v_{k,e}~~~,
 \end{equation}
 with $\delta v_{k; i,e}$ the respective contributions from the inelastic and elastic scatterers computed with
 $\rho_{i,e}=1$,
 the likelihood function is a positive semi-definite quadratic form in these two parameters.  Hence for fixed values of
 the other six parameters $\psi_{i,e},\,R_{i,e},\,D_{i,e}$, the minimization of $\chi^2$ with respect to the
 parameters $\rho_{i,e}$ can be accomplished algebraically by solving a pair of linear equations
 in the  two variables $\rho_{i,e}$, as described in Section 3 of \cite{adler2} (Section III of the arXiv version).

As a consequence, it is only necessary to  search numerically a six parameter space.  We performed the fit by doing a survey of the six parameter space on a coarse mesh to
find a good starting point for the minimization program Minuit, then doing a finer mesh survey centered on the minimum found by Minuit, giving an improved starting point for
another application of Minuit, and so forth, until this process converged.  The results are given in Tables III through VII.  Tables III and IV give the input experimental
values of the velocity anomalies and their estimated errors, together with the fits obtained by our search program (using an adaptive integration method). The chi-squared
value at the minimum was 2.7, and the values of the  eight parameters at the chi-squared
minimum are given in Table V.  Using these parameters, we also recalculated the theoretical velocity anomalies using a 200,001 point trapezoidal integration, as well as printing out
the separate contributions of the elastic and inelastic shells to the total theoretical value, as given in Tables VI and VII.  We see that the total theoretical anomaly is approximately
independent of the integration method, and that the elastic and inelastic shells both make important contributions in our model.

In about two years time, the Juno flyby should give an additional data point; the prediction of our fit for this flyby is a large anomaly of 11.6 mm/s, as also shown in
Table IV.

\begin{table} [t]\label{table:fit1}
\centering
\caption{Experimental values and 11 flyby fit}
\begin{tabular} {|c| c |c| c| c| c | c|}
\hline\hline
~~~&~~~GLL-I~~~ & ~~~GLL-II~~~ & ~~~NEAR~~~ & ~~~Cassini~~~ & ~~~Rosetta~~~ & ~~~Messenger~~~ \\
\hline
$\delta_{v;A}$ (mm/s) & 4.48 & -4.60 & 13.46 &-1.02  &1.80  &0.02 \\
$\sigma_A$    &0.3 &1.0  &0.01  &1.0  &0.03  &0.01  \\
$\delta_{v;th}$ (mm/s)&4.09  &-4.66  &13.5  &-0.807  &1.80  &0.01  \\

\hline
\end{tabular}
\end{table}

\begin{table} \label{table:fit2}
\centering
\caption{Experimental values and 11 flyby fit -- continued, and prediction for the future Juno flyby}
\begin{tabular} {|c| c |c| c| c| c ||c|}
\hline\hline
~~~&~~~Rosetta II~~~ & ~~~Rosetta III~~~ & ~~~EPOXI I~~~ & ~~~EPOXI II~~~ & ~~~EPOXI III~~~&~~~Juno~~~ \\
\hline
$\delta_{v;A}$ (mm/s) &0.0  &0.0  &0.0 &0.0 &0.0 &--- \\
$\sigma_A$    &0.1  &0.1  &0.1  &0.1  &0.1&---   \\
$\delta_{v;th}$ (mm/s)&0.0  &0.006  &0.0  &0.0  &0.0&11.6 \\

\hline
\end{tabular}
\end{table}

\begin{table}[b] \label{table:fitparam}
\centering
\caption{Parameter values for 11 flyby fit}
\begin{tabular} {|c| c |c| c| c| c |c|c|}
\hline\hline
$10^7 \times \rho_i$~(${\rm km}$)~~& $10^2 \times \rho_e$~(${\rm km}$)~~&$\psi_i$~(rad)~~&$\psi_e$~(rad)~~&$R_i$~(${\rm km}$)~~&$D_i$~(${\rm km}$)~~&$R_e$~(${\rm km}$)~~&$D_e$~(${\rm km}$)~~\\
\hline
0.398  & 0.272  & 2.79  & 0.0603  & 7561  & 2038 & 12526  & 1668  \\

\hline
\end{tabular}
\end{table}

\begin{table} [t]\label{table:fit3}
\centering
\caption{11 flyby fit (tot) and contributions from inelastic (i) and elastic (e) shells}
\begin{tabular} {|c| c |c| c| c| c | c|}
\hline\hline
~~~&~~~GLL-I~~~ & ~~~GLL-II~~~ & ~~~NEAR~~~ & ~~~Cassini~~~ & ~~~Rosetta~~~ & ~~~Messenger~~~ \\
\hline

$\delta_{v;th~tot}$ (mm/s)&3.97  &-4.65 &13.4  &-0.755  &1.76  &0.010  \\
$\delta_{v;th~i}$ (mm/s)&8.11  &7.63  &13.5  &14.3  &8.98  &0.010  \\
$\delta_{v;th~e}$ (mm/s)&-4.14  &-12.3  &-0.03  &-15.1  & -7.22 &0.00  \\
\hline
\end{tabular}
\end{table}

\begin{table} [b]\label{table:fit4}
\centering
\caption{11 flyby fit (tot) and contributions from inelastic (i) and elastic (e) shells -- continued}
\begin{tabular} {|c| c |c| c| c| c |}
\hline\hline
~~~&~~~Rosetta II~~~ & ~~~Rosetta III~~~ & ~~~EPOXI I~~~ & ~~~EPOXI II~~~ & ~~~EPOXI III~~~ \\
\hline

$\delta_{v;th~tot}$ (mm/s)&0.0  &0.008  &0.0  &0.0  &0.0 \\
$\delta_{v;th~i}$ (mm/s)&0.0  &5.14  &0.0  &0.0  &0.0 \\
$\delta_{v;th~e}$ (mm/s)&0.0  &-5.13  &0.0  &0.0  &0.0 \\

\hline
\end{tabular}
\end{table}

\vfill\break

\section{Plots of the inelastic and elastic shell acceleration along the trajectory, and of the cumulative velocity anomaly}

In this section  we describe sample plots, given at the end of the paper,  of the spacecraft downtrack acceleration from the inelastic and elastic dark matter shells (i.e., the inelastic and elastic shell  accelerations  projected along the spacecraft trajectory) and the cumulative
asymptotic velocity anomaly $\delta V_{\infty}$ (obtained by integrating the total downtrack acceleration with respect to time, dividing by the velocity at infinity, and multiplying by $10^6$ to get
an answer in ${\rm mm/s}$ rather than in ${\rm km/s}$).\footnote{ Note that the cumulative asymptotic velocity anomalies exhibited are inferred from the corresponding {\it kinetic energy} anomalies
in the asymptotic region where the potential energy is negligible; these graphs {\it do not} give the changes in the along-orbit velocity in sub-asymptotic regions where the potential energy must be taken into account.  See Appendix A for a detailed discussion.}
Plots are given for the five flybys GLL-I, GLL-II, NEAR, Cassini, and Rosetta which had
significant nonzero velocity anomalies.  For NEAR, we also plot the downtrack, crosstrack, and normal total accelerations, defined respectively as the inner product of the
total acceleration with $(dx_o/dt,\,dy_o/dt,\,0)/v$, $(-dy_o/dt,\,dx_o/dt,\,0)/v$, and $(0,0,1)$, with $v=\big((dx_o/dt)^2+(dy_o/dt)^2\big)^{1/2}$ the downtrack velocity.  We also give  plots of the cumulative asymptotic velocity anomaly and the downtrack acceleration from the inelastic and elastic shells for the future Juno flyby, for which a large anomaly is expected based on our fit.   Because of the integrable singularities in the accelerations arising from the Jacobian that is discussed
in detail in \cite{adler2}, we have plotted the accelerations on a semi-log rather than a linear plot; on a linear plot, only a spike at the Jacobian peaks appears in any detail.
These plots may facilitate comparison of the dark matter model with other anomalous force models devised to explain the flyby anomalies; on request by email to adler@ias.edu, we
can also furnish the numerical tables from which these plots were made.

\section{Predictions for flyby orientation anomalies}
Using the formulas given in Appendix A, we can calculate the anomalies in the angular momentum vector per unit mass $\vec L$,
and the Laplace--Runge--Lenz vector per unit mass squared $\vec A$, produced by passage of a flyby through the dark matter
shells.  Together with the change in the energy, or equivalently the change in the along-trajectory asymptotic velocity,
on which we have based the fit to our model, these additional anomalies completely characterize the final asymptotic state
of the flyby.  The results of this calculation are given in Table VIII.

\begin{table}[b] \label{table:LandA}
\centering
\caption{Anomalies in $\delta \vec L(\infty)$ in ${\rm km}^2/{\rm s}$, and in $\delta \vec A(\infty)$ in ${\rm km}^3/{\rm s}^2$, with components given
on the $x$,$y$, and normal $z$ axes of the flyby plane.   }
\begin{tabular} {|c|| c |c| c| c| c |c|}
\hline\hline
 flyby &$\delta L_x$ &$\delta L_y$ & $\delta L_z$ &$\delta A_x$& $\delta A_y$&$\delta A_z$\\
\hline

 GLL-I &  0.13E-01& -0.23E-01&  0.12E-01&  0.47E+00& -0.17E+00& -0.12E+00\\
 GLL-II & -0.32E-01& -0.27E-01& -0.47E-01& -0.77E+00&  0.65E+00&  0.31E+00\\
 NEAR &  0.19E-01& -0.31E-01&  0.52E-01&  0.13E+01&  0.20E+00& -0.16E+00\\
 Cassini & -0.29E-01& -0.22E-01&  0.26E-01&  0.29E+00& -0.42E+00&  0.47E+00\\
 Rosetta &  0.96E-02& -0.25E-01&  0.20E-02&  0.13E+00& -0.16E+00& -0.57E-01\\
 Messenger &  0.00E+00&  0.00E+00&  0.00E+00&  0.00E+00&  0.00E+00&  0.00E+00\\
 Rosetta II&  0.00E+00&  0.00E+00&  0.00E+00&  0.00E+00&  0.00E+00&  0.00E+00\\
 Rosetta III& -0.14E-02&  0.14E-01& -0.31E-02& -0.25E-01& -0.12E+00&  0.14E-01\\
 EPOXI I& -0.40E-26&  0.56E-24&  0.12E-22&  0.16E-21& -0.16E-23&  0.17E-25\\
EPOXI II&  0.00E+00&  0.00E+00&  0.00E+00&  0.00E+00&  0.00E+00&  0.00E+00\\
EPOXI III &  0.00E+00&  0.00E+00&  0.00E+00&  0.00E+00&  0.00E+00&  0.00E+00\\
Juno & -0.14E-01&  0.19E-01&  0.64E-01&  0.18E+01&  0.36E+00&  0.16E+00\\
\hline
\end{tabular}
\end{table}
\vfill\eject

\section{Formulas for the flyby temperature change, and the fraction of scattered nucleons}

In addition to flyby velocity changes arising from the average over the scattering cross section of the collision-induced velocity change, there will be spacecraft
temperature increases arising from the mean squared fluctuation of the collision-induced velocity change.  These are estimated in Appendix B (which simplifies an earlier
account given in \cite{adler4}), giving for inelastic scattering the formula
\begin{equation}\label{eq:esti2}
T_f-T_i
\sim  \frac{ 0.13 {}^{\circ}{\rm K} }{|\langle \cos \theta \rangle|}   \frac{\sqrt{\Delta m m_2'} c^2}{\rm MeV}~~~,
\end{equation}
with $m_2'$ the dark matter secondary mass, with $\Delta m=m_2-m_2'$ the dark matter exothermic energy release, and with $\theta$ the nucleon scattering angle.
Additionally, taking  the ratio of the flyby velocity change $\sim 1 \,{\rm cm}/{\rm s} \sim 0.3 \times 10^{-10}c$ to Eq. \eqref{eq:velchi} of Appendix B for the nucleon velocity change in a single inelastic scatter, we see that the fraction $F$ of flyby nucleons undergoing inelastic scatters is of order
\begin{equation}\label{eq:frac}
F \sim \frac{0.2 \times 10^{-10}}{|\langle \cos \theta \rangle|} \frac{m_1}{\sqrt{\Delta m m_2'}}~~~,
\end{equation}
with $m_1$ the nucleon mass.  These formulas can be used to place constraints on the model, when upper bounds
on  flyby temperature changes, and on possible radiation damage to flyby electronic components, are available.

\section{Predictions for orbit radius changes of earth-orbiting satellites}

The formulas of Eq. \eqref{eq:param} also describe closed elliptical orbits when $e<1$, with the semi-major axis $a$ given by $a=p/(1-e^2)$.  So the same program used to evaluate
the flyby anomalies can also be used to evaluate the secular change $\delta a/a$ for satellites in closed orbits, induced by scattering on the dark matter shells.  To get the secular
change over a single orbit,
we take the  limits of the  $\theta_o$ integration for the energy change $\delta E$  as  $-\pi <\theta_o \leq \pi$, and use  the formula
\begin{equation}
\frac{\delta a}{a}=\frac{2 a \delta E} {GM_{\oplus}}~~~.
\end{equation}
Sample results of this calculation, for circular orbits ($e$=0) and with the result re-scaled to give the secular change over a year, are given
in Table IX.  We see that the secular changes depend strongly on the orbit, and can range from very small to large.  For polar orbits, they are relatively small, and for orbits
beyond a radius of 20,000 km they are negligible.  This table shows that it will be important to impose closed orbit constraints on the model.  The data needed for this (for
satellites beyond an altitude of 303 km) are the orbit semi-major axis $a$, the ellipticity $e$, and the earth axis polar angle $I$ and azimuthal angle $\alpha$ on the satellite orbit plane, together with the observed value of, or bound on, the annual anomalous increment $\delta a/a$.

Added note:  The above paragraph is what appeared in the first version of this paper.  As a result of our posting this to arXiv, Edward Wright sent us several emails with
detailed data for COBE, GP-B and other satellites.  The secular changes in orbit radius for these satellites are orders of magnitude smaller than predicted by Table IX.
For example, for COBE, which is in a near-polar orbit with an inclination of 99 degrees, and a=7278 km, the observed $d\ln(a)/dt=-0.00009/{\rm yr}$, compared with a
prediction of order $0.02/{\rm yr}$ from Table IX.  And for GP-B, which is in an almost exactly polar orbit with inclination of 90.007 degrees, and a=7027.4 km, quoting from Wright's email, ``from 2004.7 to 2004.8 GP-B was in a ``drag-free'' mode, where the spacecraft
tracked the motion of one (of) the rotors that was shielded from atmospheric drag and radiation pressure. I have attached a plot of the orbital rate of GP-B vs date. The
drag-free epoch is quite obvious....during the drag-free epoch $|d\ln(a)/dt|$ is clearly less than 0.000005/yr, while your model predicts about 0.028/yr.''  We agree with Wright
that this data rules out the model analyzed in this paper.  Any further modeling of the flyby anomalies must, as an essential component, include closed orbit satellite
constraints in the fitting procedure.  Our perspective is that this should wait until after the Juno flyby, which has an orbit similar enough to that of GLL-I (compare Figs. 1-3 for
GLL-I with Figs. 19-21 for Juno) to have the potential, independent of modeling,  to confirm or rule out the presence of a kinetic energy anomaly.  If the Juno flyby can be
tracked through earth approach, and shows an anomaly, analysis of the tracking data using the methods of Appendix A would give important information about the radius range where the
anomaly arises.

\section{Discussion}
In \cite{adler2} we gave an estimate for the total mass $M_e$ and $M_i$  in the elastic and inelastic dark matter shells,
 \begin{align}\label{eq:mass}
 M_e\simeq &4 \pi^{5/2} \rho_eD_e m_1/\sigma_{\rm el}~~~,\cr
 M_i \simeq & 4 \pi^{5/2} \rho_iD_i m_1/(B_{\rm inel}\sqrt{2 \Delta m/m_2'}~)~~~,\cr
 \end{align}
 where in $M_i$ we have allowed for the possibility that $\Delta m$ is much smaller than $m_2'$  (rather than
 $\Delta m \sim m_2'$, as assumed in \cite{adler2}).
 Here $\sigma_{\rm el}$ is the threshold cross section for the scattering of the elastic dark matter population
 on nucleons,  $B_{\rm inel}$ is the coefficient of the $\cos \theta$ term in the near threshold cross section
 for exothermic scattering of the inelastic dark matter population on nucleons (see Eq. (7) of \cite{adler2}),
 and $m_1$, $m_2'$, and $\Delta m$ are as defined in Section VI and Appendix B.
 Using these estimates, and the upper bound \cite{adler3}  of
  $4 \times 10^{-9} M_{\oplus} \sim 1.4 \times 10^{43} {\rm GeV}/c^2$ on the mass of dark matter in orbit around the earth
 between the 12,300 km radius of the LAGEOS satellite orbit and the moon's orbit, one can get
 lower bounds on $\sigma_{\rm el}$ and $B_{\rm inel}$.  For example, referring to Table V, from  $\rho_eD_e=4.537\, {\rm km}^2$  we get
\begin{equation}\label{eq:bound1}
\sigma_{\rm el} \geq 2.2\times 10^{-31} {\rm cm}^2~~~,
\end{equation}
while from $\rho_iD_i=0.0000811\,{\rm km}^2$, taking into account the fact that only a fraction
\begin{equation}\frac{1}{\surd \pi} \int_{L}^{\infty} dz \exp(-z^2) \simeq 0.50 \times 10^{-3}
\end{equation}
\big(with $L=(12,300-7561)/2038=2.325$\big) of the inelastic dark matter shell lies above the LAGEOS orbit, we correspondingly get
 \begin{equation}\label{eq:bound2}
B_{\rm inel}\sqrt{\Delta m/m_2'}   \geq 1.4 \times 10^{-39} {\rm cm}^2~~~.
\end{equation}
These bounds are consistent with the cross section range arrived at from various constraints in \cite{adler1}.  In comparing these bounds with cross section limits derived
from dark matter direct detection experiments, two points should be kept in mind.  The first is that if the dark matter mass is much lower than a GeV, the recoils from
elastic scattering on nucleons of galactic halo dark matter will be below the threshold for detection in direct searches, and so no useful constraints on the elastic dark matter nucleon
scattering cross section are
obtained.  The second is that with two species of dark matter, as in our model, the fraction $f_i$ of the galactic halo consisting of inelastically scattering dark matter
can be much smaller than unity.  In this case a direct detection cross section bound $\sigma_{\rm MAX}$, obtained assuming a single species of dark matter in the
galactic halo, implies a significantly larger bound $\sigma_{\rm MAX}/f_i$ for the dark matter-nucleon scattering cross section of the inelastic component.

As opposed to the fits given in \cite{adler2}, the current fit to 11 flybys places the elastic and inelastic dark matter
shells much closer to earth, well within the orbits of geostationary and Global Positioning System satellites.  So
the effect on these satellites should be minimal, but it now becomes important to examine the implications of our model for
satellites that are in low or medium earth orbit, as initiated  in Section VII and Table IX. (See in this regard the Added note in Section VII.)    It will also be important to evaluate  implications of a low earth
orbit exothermic scattering dark matter shell for earth's heat balance.  We pose these as significant questions to be
addressed in future investigations.
\section{Acknowledgments}

I wish to thank James K. Campbell  for email correspondence furnishing me with the parameters for the
five new flybys Rosetta II, Rosetta III, EPOXI I, EPOXI II, and EPOXI III, for suggesting that I calculate the
plots shown in the figures to facilitate comparison with other models, and for comments on the draft of
this paper.    I also wish to thank John D. Anderson for his assistance
in supplying the parameters needed for my calculation, and to thank Edward Wright for sending data on the secular radius
changes of COBE, GP-B, and other satellites, after the first version of this paper was posted to arXiv.   I  wish to thank Institute for Advanced
Study Members Kfir Blum, David Spiegel, and Tiberiu Tesileanu for helpful advice on plotting the figures, and Freeman Dyson for asking
about the dark matter mass.  My work has
been partially supported by the Department of Energy under grant no DE-FG02-90ER40542.

\appendix
\section{ Calculation of energy, angular momentum, and Laplace--Runge--Lenz vector anomalies, and their relation to local velocity and position anomalies}

When a perturbation of force per unit mass $\delta \vec F$ is applied to a flyby spacecraft,
the total change in energy per unit mass at time $t$ is given by the work integral
\begin{equation}
\delta E(t)=\int_{-\infty}^t du \frac {dE(u)}{du} = \int_{-\infty}^t du\, \vec v_o(u) \cdot \delta \vec F(u)=\int_{-\infty}^t du\, \frac{ d\vec x_o(u)}{du} \cdot \delta \vec F(u)~~~,
\end{equation}
which when divided by the velocity at infinity and multiplied by $10^6$ is the quantity plotted as the
``cumulative asymptotic velocity anomaly $\delta V_{\infty}$'' in the figures.  Taking a first variation of the energy
per unit mass (where we write $r_o(t)=|\vec x_o(t)|$ and $\mu=G M_{\oplus}$),
\begin{equation}
E=\frac{1}{2} \vec v_o(t)^2 - \frac{\mu}{r_o(t)}
\end{equation}
we get
\begin{align}
\delta E(t) =& \int_{-\infty}^t du\, \frac{ d\vec x_o(u)}{du} \cdot \delta \vec F(u)\cr
= &\delta \vec v_o(t) \cdot \vec v_o(t)
+ \frac{\mu \delta \vec x_o(t) \cdot \vec x_o(t)}{r_o(t)^3}~~~.\cr
\end{align}
This formula gives the relation between the cumulative energy anomaly at time $t$, and the position and velocity anomalies
of the orbit $\delta \vec x_o(t)$ and $\delta \vec v_o(t)$ at time $t$.  At time $t=\infty$, the potential energy
contribution vanishes, giving a formula relating the work integral taken from $t=-\infty$ to $t=\infty$ to the asymptotic
along-track velocity anomaly $\delta V_{\infty}$, which as this derivation makes clear is an asymptotic flyby
energy anomaly.

In addition to the energy, a Kepler orbit has conserved angular momentum vector per unit mass $\vec L$,
\begin{equation}
\vec L = \vec x_o(t) \times \vec v_o(t)~~~,
\end{equation}
and conserved Laplace--Runge--Lenz vector per unit mass squared $\vec A$,
\begin{equation}
\vec A=\vec v_o(t) \times \vec L - \frac{\mu \vec x_o(t)}{r_o(t)}~~~.
\end{equation}

Taking the time derivatives of these quantities, substituting the perturbed equation of motion
\begin{equation}
\frac{d \vec v_o(t)}{dt}= -\frac{\mu \vec x_o(t)}{r_o(t)^3}+\delta \vec F(t)~~~,
\end{equation}
and then integrating with respect to time, and equating these to first variations of the conserved quantities, we get formulas
\begin{align}
\delta \vec L(t)=&\int_{-\infty}^{t} du \,\vec x_o(u) \times  \delta \vec F(u)\cr
 =&\delta \vec x_o(t) \times \vec v_o(t)+ \vec x_o(t) \times  \delta \vec v_o(t)~~~,\cr
\end{align}
and
\begin{align}
\delta \vec A(t)=&\int_{-\infty}^t du [2\delta \vec F(u) \cdot \vec v_o(u) \vec x_o(u) - \delta \vec F(u) \cdot \vec x_o(u) \vec v_o(u)
-\vec x_o(u) \cdot \vec v_o(u) \delta \vec F(u)] \cr
=&2\vec v_o(t) \cdot \delta \vec v_o(t) \vec x_o(t) + \vec v_o(t)^2 \delta \vec x_o(t) -\delta \vec v_o(t) \cdot \vec x_o(t) \vec v_o(t)
-\vec v_o(t) \cdot \delta \vec x_o(t) \vec v_o(t)\cr
 -&\vec v_o(t) \cdot \vec x_o(t) \delta \vec v_o(t)
-\frac{\mu \delta \vec x_o(t)}{r_o(t)} + \frac{\mu \vec x_o(t) \vec x_o(t) \cdot \delta \vec x_o(t)}{r_o(t)^3} ~~~.\cr
\end{align}
Setting $t=\infty$ in these formulas gives, in terms of integrals over the unperturbed orbit of the force perturbation $\delta \vec F$, an expression
for the asymptotic orbit orientation perturbations $\delta \vec L(\infty)$ and $\delta \vec A(\infty)$. These in turn can be calculated
using the above formulas from the observed asymptotic perturbations $\delta \vec x_o(\infty)$ and $\delta \vec v_o(\infty)$ obtained from
tracking the flyby position and velocity,
\begin{equation}
\delta \vec L(\infty) =\delta \vec x_o(\infty) \times \vec v_o(\infty)+ \vec x_o(t\infty) \times  \delta \vec v_o(\infty)~~~,
\end{equation}
and
\begin{align}
\delta \vec A(\infty)=&
2\vec v_o(\infty) \cdot \delta \vec v_o(\infty) \vec x_o(\infty ) + \vec v_o(\infty)^2 \delta \vec x_o(\infty) -\delta \vec v_o(\infty) \cdot \vec x_o(\infty) \vec v_o(\infty)\cr
-&\vec v_o(\infty) \cdot \delta \vec x_o(\infty) \vec v_o(\infty) -\vec v_o(\infty) \cdot \vec x_o(\infty) \delta \vec v_o(\infty)~~~. \cr
\end{align}
As a point of consistency, we note that the same values of $\delta \vec L(\infty)$ and $\delta \vec A(\infty)$ are obtained irrespective of where on the
asymptotic outgoing trajectory these formulas are evaluated.  If after a first evaluation,  they are then evaluated at a time $T$ later on, the quantities
$\vec x_o(\infty)$ and $\delta \vec x_o(\infty)$ are augmented respectively by $\vec v_o(\infty) T$ and $\delta \vec v_o(\infty) T$.  One can then check that
the additions proportional to $T$ cancel out of the equations for $\delta \vec L(\infty)$ and $\delta \vec A(\infty)$.

\section{ Temperature change arising from velocity fluctuations}

In \cite{adler1} we considered the velocity change when a spacecraft nucleon of
mass $m_1\simeq 1 {\rm GeV}$ and initial velocity $\vec u_1$
scatters from a dark matter particle of mass $m_2$ and initial
velocity $\vec u_2$, into an outgoing nucleon of mass $m_1$ and
velocity $\vec v_1$, and an outgoing secondary dark matter
particle of mass $m_2'=m_2-\Delta m$ and velocity $\vec v_2$ . (In
the elastic scattering case, one has $m_2'=m_2$ and $\Delta m=0$.)
Under the assumption that  both initial particles are nonrelativistic, so that
$|\vec u_1|<<c, |\vec u_2|<<c$,  a
straightforward calculation shows that the outgoing nucleon velocity is given by
\begin{equation}\label{eq:vel1}
 \vec v_1=\frac {m_1 \vec u_1 + m_2 \vec
u_2}{m_1+m_2'}+w\hat v_{\rm out}~~~.
\end{equation}
Here $w>0$ is given\footnote{The notation $t$ was used in \cite{adler1} for what we here term $w$;
the change in notation avoids confusion with use of $t$ for time.} by taking the square root of
\begin{equation}\label{eq:tdef}
w^2=\frac{m_2m_2'}{(m_1+m_2)(m_1+m_2')}(\vec u_1-\vec u_2)^2 +
\frac{\Delta m~ m_2'}{m_1(m_1+m_2')} \Big[2 c^2 - \frac{(m_1\vec
u_1+m_2\vec u_2)^2}{(m_1+m_2)(m_1+m_2')}\Big]~~~,
\end{equation}
and $\hat v_{\rm out}$ is a kinematically free unit vector.  Denoting
by $\theta$ the angle between $\hat v_{\rm out}$ and the entrance channel
center of mass nucleon velocity $\vec u_1-(m_1 \vec u_1 +m_2 \vec u_2)/(m_1+m_2)=m_2(\vec u_1-\vec u_2)/(m_1+m_2)$,
and assuming that the center of mass scattering amplitude is a function $f(\theta)$ only of
this polar angle, the average over scattering angles of the outgoing nucleon velocity is given by
\begin{equation}\label{eq:vel2}
 \langle \vec v_1\rangle =\frac {m_1 \vec u_1 + m_2 \vec
u_2}{m_1+m_2'}+w \langle \cos \theta \rangle \frac{\vec u_1-\vec u_2}{|\vec u_1-\vec u_2|}~~~,
\end{equation}
 with $\langle \cos \theta \rangle$ given by
\begin{equation}\label{eq:costhetdef}
 \langle \cos \theta \rangle = \frac {\int_0^{\pi}d\theta \sin
\theta \cos \theta |f(\theta)|^2} {\int_0^{\pi}d\theta \sin \theta
|f(\theta)|^2}~~~.
\end{equation}
Subtracting $\vec u_1$ from Eq. \eqref{eq:vel2} gives the formula
for the average velocity change used in \cite{adler1} and \cite{adler2} to calculate the
flyby velocity change,
\begin{equation}\label{eq:vel3}
\langle \delta \vec v_1\rangle =\frac {m_2 \vec u_2 - m_2' \vec
u_1}{m_1+m_2'}+w\langle \cos \theta \rangle \frac {\vec u_1-\vec
u_2}{|\vec u_1-\vec u_2|}~~~.
\end{equation}

However, in addition to contributing to an average change in the outgoing nucleon velocity, dark
matter scattering will give rise to fluctuations in this velocity, which have a mean square magnitude
given by
\begin{equation}\label{eq:fluct1}
\langle (\vec v_1-\langle \vec v_1 \rangle)^2 \rangle
=w^2 \langle \Big(\hat v_{\rm out}-\langle \cos \theta \rangle  \frac{\vec u_1-\vec u_2}{|\vec u_1-\vec u_2|}\Big)^2 \rangle
=w^2 (1-\langle \cos \theta \rangle ^2)~~~.
\end{equation}
This fluctuating velocity leads to an average temperature increase of the nucleon, per single scattering, of order
\begin{equation}\label{eq:tempincr1}
\langle \delta T \rangle \sim\frac{2}{3}  \frac {m_1}{2k_B} \langle (\vec v_1-\langle \vec v_1 \rangle)^2 \rangle
=\frac {m_1} {3k_B} w^2 (1-\langle \cos \theta \rangle ^2)~~~,
\end{equation}
with $k_B$ the Boltzmann constant.
In analogy with the treatment of the velocity change $\delta \vec v_1$ in  \cite{adler1}, to calculate $dT/dt$, the time rate of change of temperature
of the spacecraft resulting from dark
matter scatters,  one multiplies the
temperature change in a single scatter $\langle \delta T\rangle $ by the number of scatters per unit time.  This latter
is given by the flux $|\vec u_1-\vec u_2|$, times the scattering
cross section $\sigma$, times the dark matter spatial and velocity
distribution $\rho\big(\vec x, \vec u_2\big)$.  Integrating out
the dark matter velocity, one thus gets for $dT/dt$ at
the  point $\vec x(t)$ on the spacecraft trajectory with velocity
$\vec u_1=d\vec x(t)/dt$,
\begin{equation}\label{eq:tdot}
dT/dt= \int d^3 u_2 \langle \delta T\rangle |\vec
u_1-\vec u_2| \sigma \rho\big(\vec x, \vec u_2\big)~~~.
\end{equation}
Integrating  from
$t_i$ to $t_f$ we get for the temperature change resulting from
dark matter collisions over the corresponding interval
of the spacecraft trajectory ,
\begin{equation}\label{eq:tchange}
T_f-T_i  =\int_{t_i}^{t_f} dt \int d^3 u_2
 \langle \delta T\rangle |\vec u_1-\vec u_2| \sigma
\rho\big(\vec x, \vec u_2\big)~~~.
\end{equation}

In the elastic scattering case, with $\Delta m=0$, $m_2'=m_2$,
the formula of Eq. \eqref{eq:tdef} simplifies to
\begin{equation}\label{eq:el}
w^2 = \left( \frac{m_2}{m_1+m_2}\right)^2(\vec
u_1-\vec u_2)^2 ~~~.
\end{equation}
In the inelastic case, as long as  $\Delta m/m_2 >> |\vec u_1-\vec u_2|^2/c^2$,  Eq. \eqref{eq:tdef} is well
approximated by
\begin{equation}\label{eq:inel}
w^2 \simeq    \Bigg( \frac{2 \Delta m ~m_2'} {
 m_1 (m_1+m_2')}\Bigg)c^2 ~~~.
\end{equation}
Since $\vec u_1$ and $\vec u_2$ are typically of order 10 ${\rm
km} ~{\rm s}^{-1}$, the temperature change in the inelastic case, per unit
scattering cross section times angular factors,  is
larger than that in the elastic case by a factor $ \sim c^2/|\vec
u_1|^2\sim 10^9$, and the restriction on $\Delta m$ is the weak condition $\Delta m/m_2 >> 10^{-9}$.

We proceed now to estimate Eq. \eqref{eq:tchange} by
 using Eqs. \eqref{eq:tempincr1},  \eqref{eq:el}, and \eqref{eq:inel}, and
making the approximations that the dark matter mass $m_2$ is much smaller than the nucleon
mass $m_1$, and that $\langle \cos \theta \rangle^2$ in Eq. \eqref{eq:tempincr1} can be neglected relative to 1.
In the elastic case, Eq. (4) of \cite{adler1} tells us that the magnitude of the
velocity change in a single collision is of order
\begin{equation}\label{eq:velche}
|\langle \delta \vec v_1 \rangle| \sim  \frac{m_2}{m_1} |\vec u_1-\vec u_2|~~~.
\end{equation}
Taking the ratio of the single collision temperature change to the single collision
velocity change, and multiplying by the flyby total velocity change $\sim 10^{-6}|\vec u_1|$,
we get as an estimate of the total temperature change
\begin{equation}\label{eq:este}
T_f-T_i \sim \frac{\delta T}{|\langle \delta \vec v_1 \rangle|}10^{-6}|\vec u_1|
\sim 10^{-6} \frac{m_2}{2k_B} |\vec u_1||\vec u_1-\vec u_2|
 \sim 0.6 \times 10^{-5}{}^{\circ}{\rm K} \Bigg(\frac{m_2 c^2}{ {\rm MeV}} \Bigg)~~~,
\end{equation}
in agreement with \cite{adler1}.

In the inelastic case, we must take into account the kinematic structure of an exothermic
inelastic differential cross section.  In the inelastic case, Eq. (5) of \cite{adler1}
tells us that the magnitude of the velocity change in a single collision is of order
\begin{equation}\label{eq:velchi}
|\langle \delta \vec v_1 \rangle| \sim\frac{\sqrt{2 \Delta m \,m_2'}}{m_1} c |\langle \cos \theta \rangle|~~~.
\end{equation}
Writing the inelastic differential cross section near threshold in the form
\begin{equation}\label{eq:siginel}
\frac{d\sigma}{d\Omega}= \frac {A_{\rm inel}}{4\pi}  k^{-1} + B_{\rm inel} \frac {3} {4\pi} \cos \theta + ...,
\end{equation}
we have
\begin{align}\label{eq:averages}
\sigma \simeq & A_{\rm inel}  k^{-1}~~~,\cr
\langle \cos \theta \rangle \simeq& B_{\rm inel} /(A_{\rm inel} k^{-1})~~~,\cr
\end{align}
with $k$ the entrance channel momentum
\begin{equation}\label{eq:kdef}
k=\frac{m_1 m_2}{m_1+m_2} |\vec u_1-\vec u_2| \simeq m_2 |\vec u_1-\vec u_2|~~~.
\end{equation}
Again taking the ratio of the single collision temperature change to the single collision
velocity change, and  multiplying by the flyby total velocity change $\sim 10^{-6}|\vec u_1|$,
with $|\vec u_1| \sim 10\,{\rm km}/{\rm s}$,
we get as an estimate of the total temperature change in the inelastic case
\begin{equation}\label{eq:esti1}
T_f-T_i \sim \frac{\delta T}{|\langle \delta \vec v_1 \rangle|}10^{-6}|\vec u_1|
\sim \frac{10^{-6}}{3k_B} \frac{1}{|\langle \cos \theta \rangle|} \frac{|\vec u_1|}{c} \sqrt{\Delta m m_2'} c^2
\sim  \frac{ 0.13 {}^{\circ}{\rm K} }{|\langle \cos \theta \rangle|} \frac{\sqrt{\Delta m m_2'} c^2}{\rm MeV}~~~.
\end{equation}
So we see that the exothermic inelastic scattering temperature rise is substantially
bigger than that from elastic scattering, as already anticipated in the remarks following Eq. \eqref{eq:inel}
above.
In particular, for the temperature change to be bounded, say, by $10  {}^{\circ}{\rm K} $, Eq. \eqref{eq:esti1} implies that
\begin{equation}
\sqrt{\Delta m m_2'} c^2 < 77 |\langle \cos \theta \rangle| {\rm MeV}~~~,
\end{equation}
which is compatible with $\sqrt{\Delta m m_2'}$ in the MeV range.  We note however, that if $\Delta m /m_2$ were much smaller than unity, then the
dark matter mass $m_2$ could be much larger than an MeV.

\begin{table}\label{table:circular}
\centering
\caption{Change $\delta a /a$ for a circular orbit in a year, versus orbit radius and earth orbit polar angle. The first column is the orbit
radius in km, with the remaining columns giving the annual $\delta a/a$ as a function of the earth polar axis angle $I$ listed at the top
of the table. For $I=0$, the orbits circulate in the same direction as the earth's rotation, and for $I=180$ the orbits circulate in the opposite
 sense as the earth's rotation. For circular orbits, the results are independent of the earth axis azimuthal angle $\alpha$. }
\begin{tabular} {|c|| c |c| c| c| c |c|c|c|c|c|}
\hline\hline
 radius&0.0 & 22.5 & 45.0 &67.5& 90.0&112.5& 135.0&157.5&180.0\\
\hline

     6878.&  0.16E+00&  0.21E+00&  0.73E-01&  0.44E-01&  0.29E-01&  0.20E-01&  0.12E-01&  0.70E-02&  0.51E-02\\
     7378.&  0.15E+00&  0.20E+00&  0.71E-01&  0.42E-01&  0.28E-01&  0.19E-01&  0.12E-01&  0.67E-02&  0.47E-02\\
     7878.&  0.13E+00&  0.18E+00&  0.61E-01&  0.37E-01&  0.24E-01&  0.16E-01&  0.10E-01&  0.56E-02&  0.31E-02\\
     8378.&  0.10E+00&  0.14E+00&  0.47E-01&  0.28E-01&  0.19E-01&  0.12E-01&  0.77E-02&  0.38E-02& -0.13E-02\\
     8878.&  0.70E-01&  0.93E-01&  0.32E-01&  0.19E-01&  0.13E-01&  0.81E-02&  0.44E-02&  0.67E-03& -0.14E-01\\
     9378.&  0.43E-01&  0.57E-01&  0.20E-01&  0.11E-01&  0.69E-02&  0.36E-02&  0.20E-03& -0.52E-02& -0.46E-01\\
     9878.&  0.24E-01&  0.31E-01&  0.10E-01&  0.52E-02&  0.18E-02& -0.15E-02& -0.61E-02& -0.17E-01& -0.12E+00\\
    10378.&  0.12E-01&  0.15E-01&  0.41E-02&  0.42E-03& -0.32E-02& -0.79E-02& -0.16E-01& -0.37E-01& -0.25E+00\\
    10878.&  0.50E-02&  0.61E-02&  0.23E-03& -0.35E-02& -0.85E-02& -0.16E-01& -0.29E-01& -0.65E-01& -0.44E+00\\
    11378.&  0.19E-02&  0.18E-02& -0.22E-02& -0.67E-02& -0.14E-01& -0.24E-01& -0.43E-01& -0.97E-01& -0.65E+00\\
    11878.&  0.65E-03& -0.71E-04& -0.35E-02& -0.88E-02& -0.17E-01& -0.30E-01& -0.54E-01& -0.12E+00& -0.80E+00\\
    12378.&  0.18E-03& -0.73E-03& -0.39E-02& -0.93E-02& -0.18E-01& -0.31E-01& -0.56E-01& -0.12E+00& -0.84E+00\\
    12878.&  0.36E-04& -0.80E-03& -0.35E-02& -0.81E-02& -0.16E-01& -0.27E-01& -0.49E-01& -0.11E+00& -0.73E+00\\
    13378.& -0.10E-05& -0.62E-03& -0.25E-02& -0.60E-02& -0.11E-01& -0.20E-01& -0.36E-01& -0.80E-01& -0.53E+00\\
    13878.& -0.61E-05& -0.39E-03& -0.16E-02& -0.37E-02& -0.70E-02& -0.12E-01& -0.22E-01& -0.49E-01& -0.33E+00\\
    14378.& -0.41E-05& -0.20E-03& -0.80E-03& -0.19E-02& -0.36E-02& -0.63E-02& -0.11E-01& -0.25E-01& -0.17E+00\\
    14878.& -0.19E-05& -0.87E-04& -0.35E-03& -0.81E-03& -0.16E-02& -0.27E-02& -0.49E-02& -0.11E-01& -0.73E-01\\
    15378.& -0.71E-06& -0.31E-04& -0.13E-03& -0.29E-03& -0.56E-03& -0.98E-03& -0.18E-02& -0.39E-02& -0.26E-01\\
    15878.& -0.22E-06& -0.95E-05& -0.38E-04& -0.89E-04& -0.17E-03& -0.30E-03& -0.53E-03& -0.12E-02& -0.80E-02\\
    16378.& -0.55E-07& -0.24E-05& -0.96E-05& -0.22E-04& -0.43E-04& -0.75E-04& -0.13E-03& -0.30E-03& -0.20E-02\\
    16878.& -0.12E-07& -0.51E-06& -0.20E-05& -0.48E-05& -0.91E-05& -0.16E-04& -0.29E-04& -0.64E-04& -0.43E-03\\
    17378.& -0.21E-08& -0.91E-07& -0.36E-06& -0.85E-06& -0.16E-05& -0.28E-05& -0.51E-05& -0.11E-04& -0.76E-04\\
    17878.& -0.31E-09& -0.14E-07& -0.54E-07& -0.13E-06& -0.24E-06& -0.42E-06& -0.75E-06& -0.17E-05& -0.11E-04\\
    18378.& -0.39E-10& -0.17E-08& -0.67E-08& -0.16E-07& -0.30E-07& -0.52E-07& -0.94E-07& -0.21E-06& -0.14E-05\\
    18878.& -0.40E-11& -0.18E-09& -0.70E-09& -0.16E-08& -0.31E-08& -0.55E-08& -0.98E-08& -0.22E-07& -0.15E-06\\
    19378.& -0.35E-12& -0.15E-10& -0.61E-10& -0.14E-09& -0.27E-09& -0.48E-09& -0.85E-09& -0.19E-08& -0.13E-07\\
    19878.& -0.26E-13& -0.11E-11& -0.44E-11& -0.10E-10& -0.20E-10& -0.35E-10& -0.62E-10& -0.14E-09& -0.93E-09\\
    20378.& -0.16E-14& -0.68E-13& -0.27E-12& -0.63E-12& -0.12E-11& -0.21E-11& -0.38E-11& -0.85E-11& -0.57E-10\\

\hline
\end{tabular}
\end{table}
\vfill\eject

\includegraphics[angle=-90]{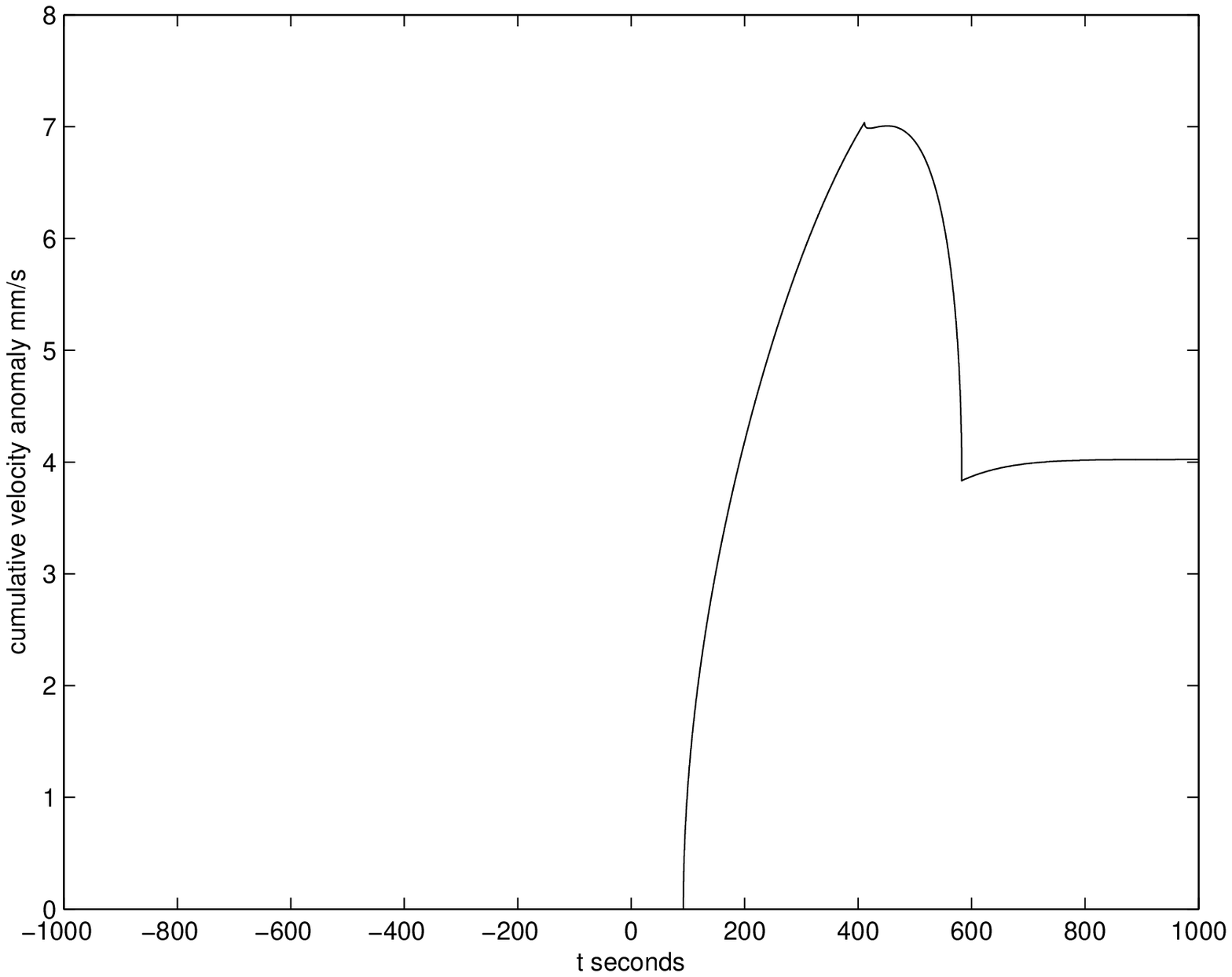}
\captionof{figure}{GLL-I cumulative asymptotic velocity anomaly $\delta V_{\infty}$ in ${\rm mm}/{\rm }s$}

\vfill\eject

\includegraphics[angle=-90]{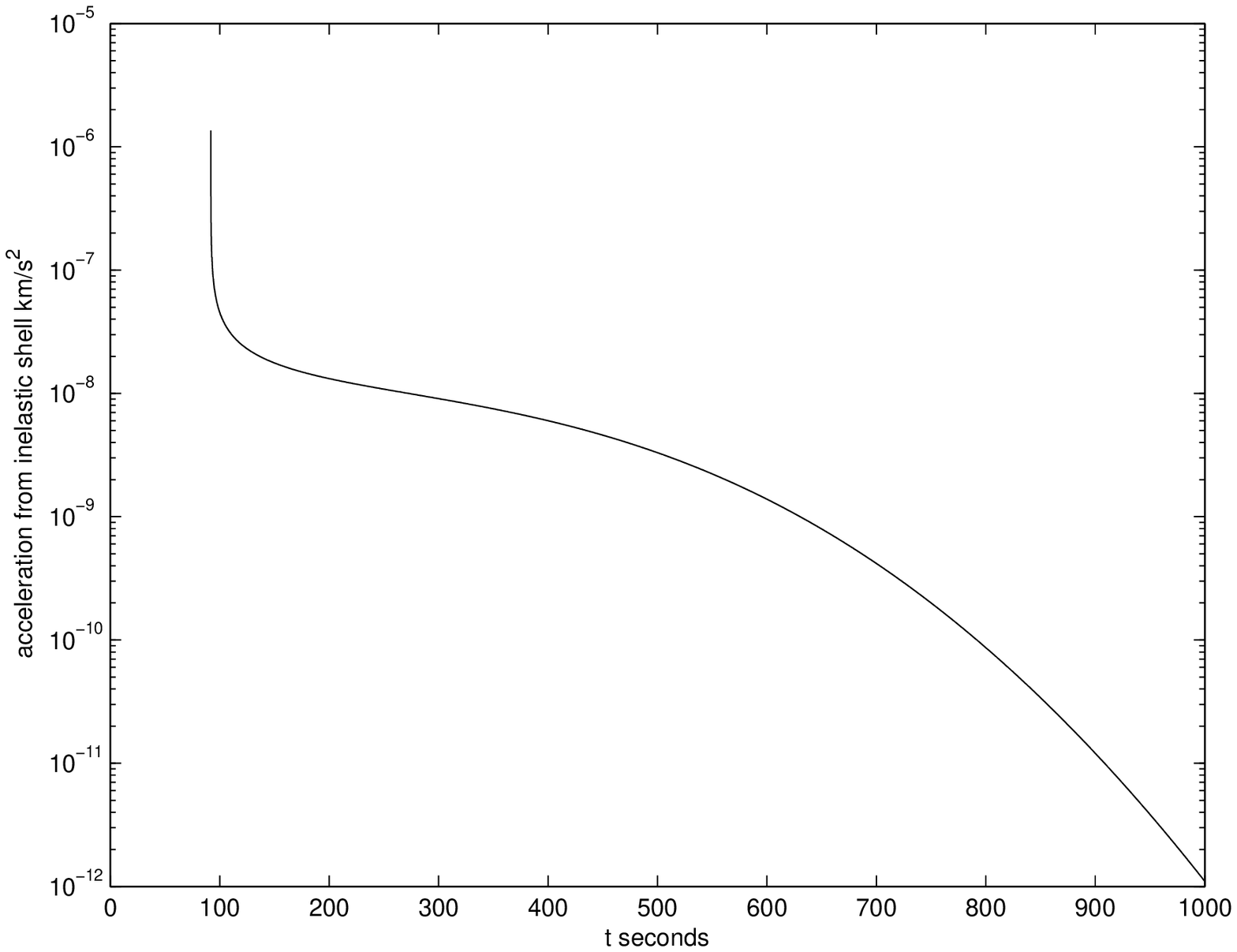}
\captionof{figure}{GLL-I downtrack acceleration from inelastic shell ${\rm km}/{\rm s}^2$ }

\vfill\eject

\includegraphics[angle=-90]{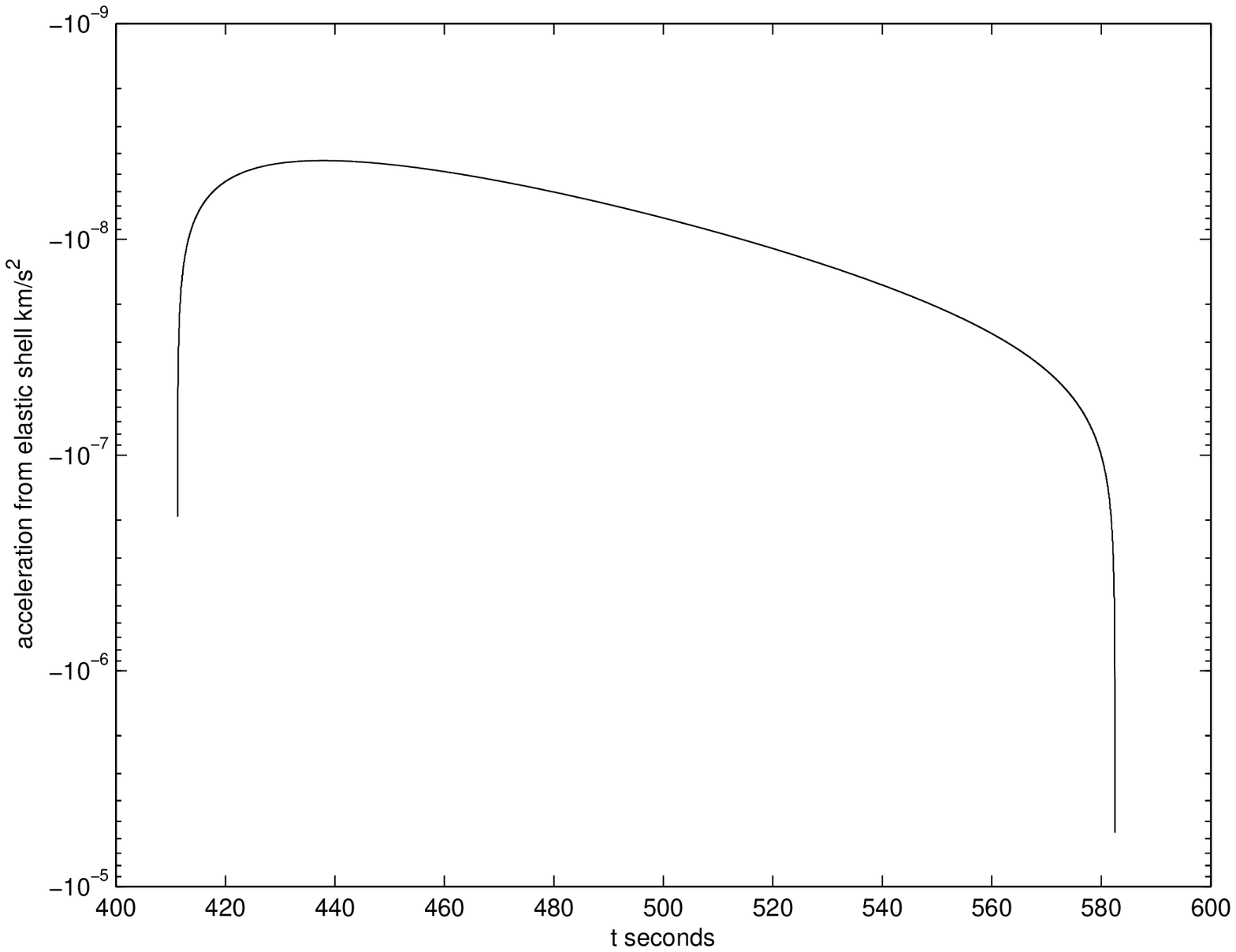}
\captionof{figure}{GLL-I downtrack acceleration from elastic shell $km/s^2$ }

\vfill\eject

\begin{figure}
\includegraphics[angle=-90]{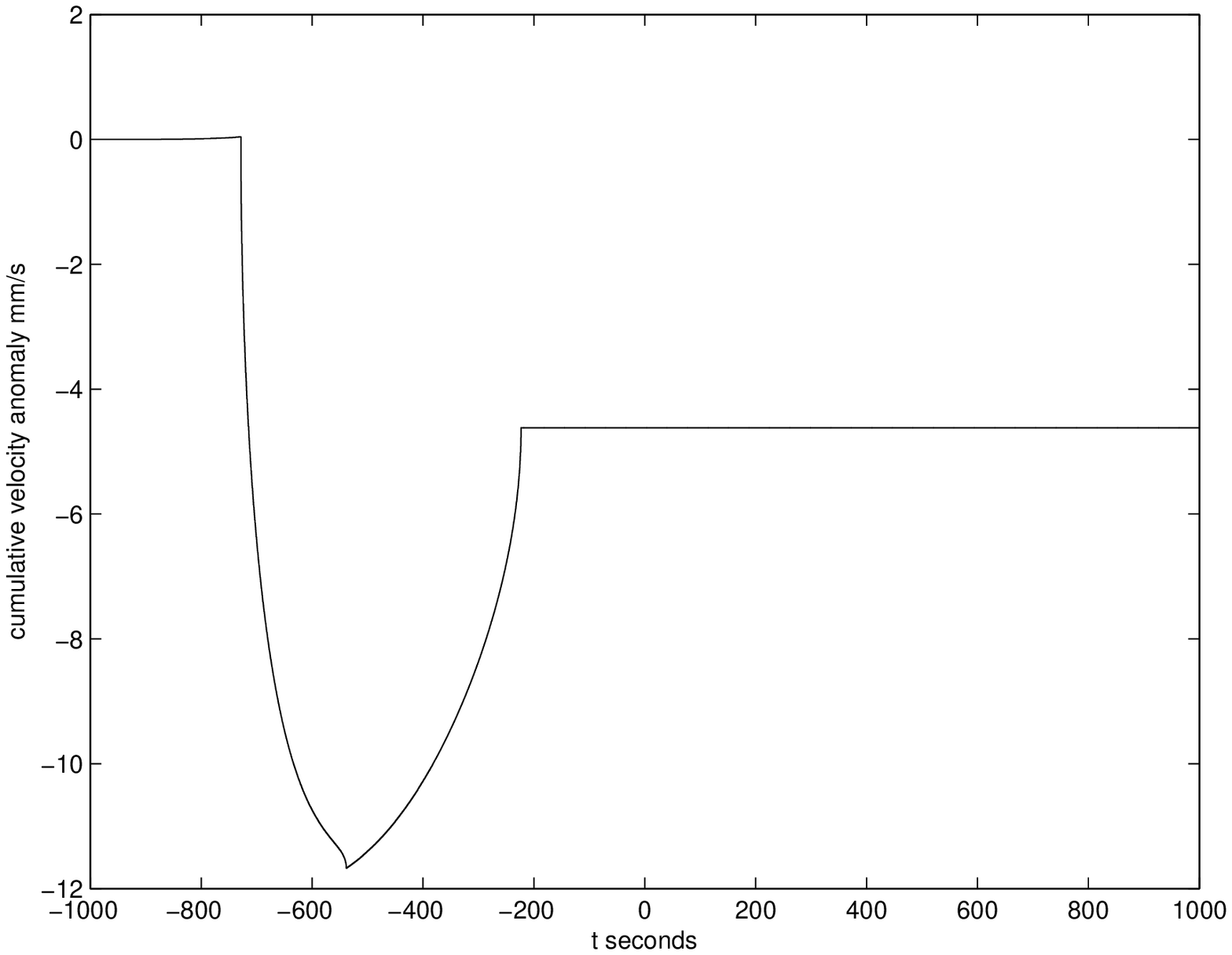}
\caption{GLL-II cumulative asymptotic velocity anomaly $\delta V_{\infty}$ in ${\rm mm}/{\rm s}$}
\end{figure}

\vfill\eject

\begin{figure}
\includegraphics[angle=-90]{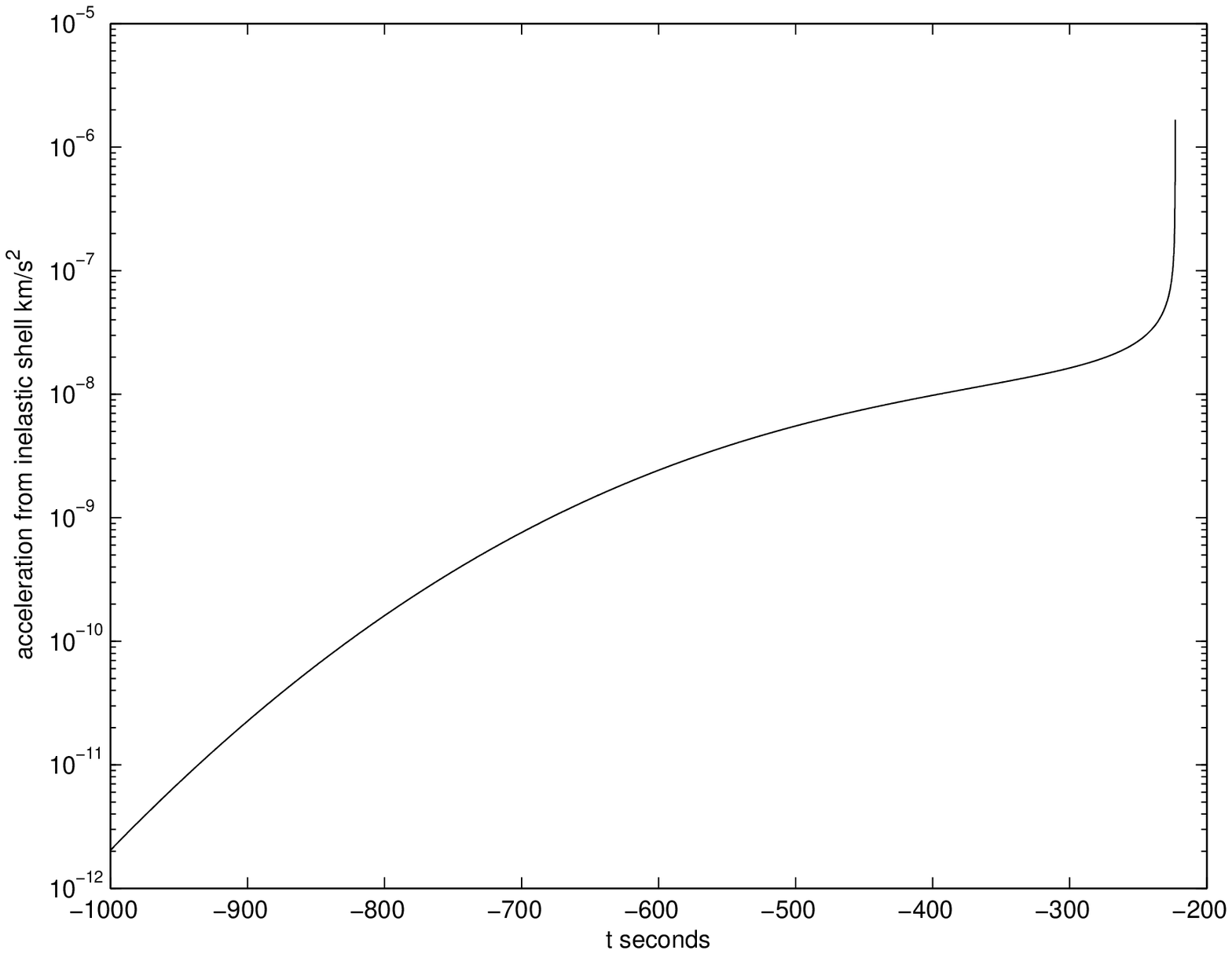}
\caption{GLL-II downtrack acceleration from inelastic shell ${\rm km}/{\rm s}^2$ }
\end{figure}

\vfill\eject

\begin{figure}
\includegraphics[angle=-90]{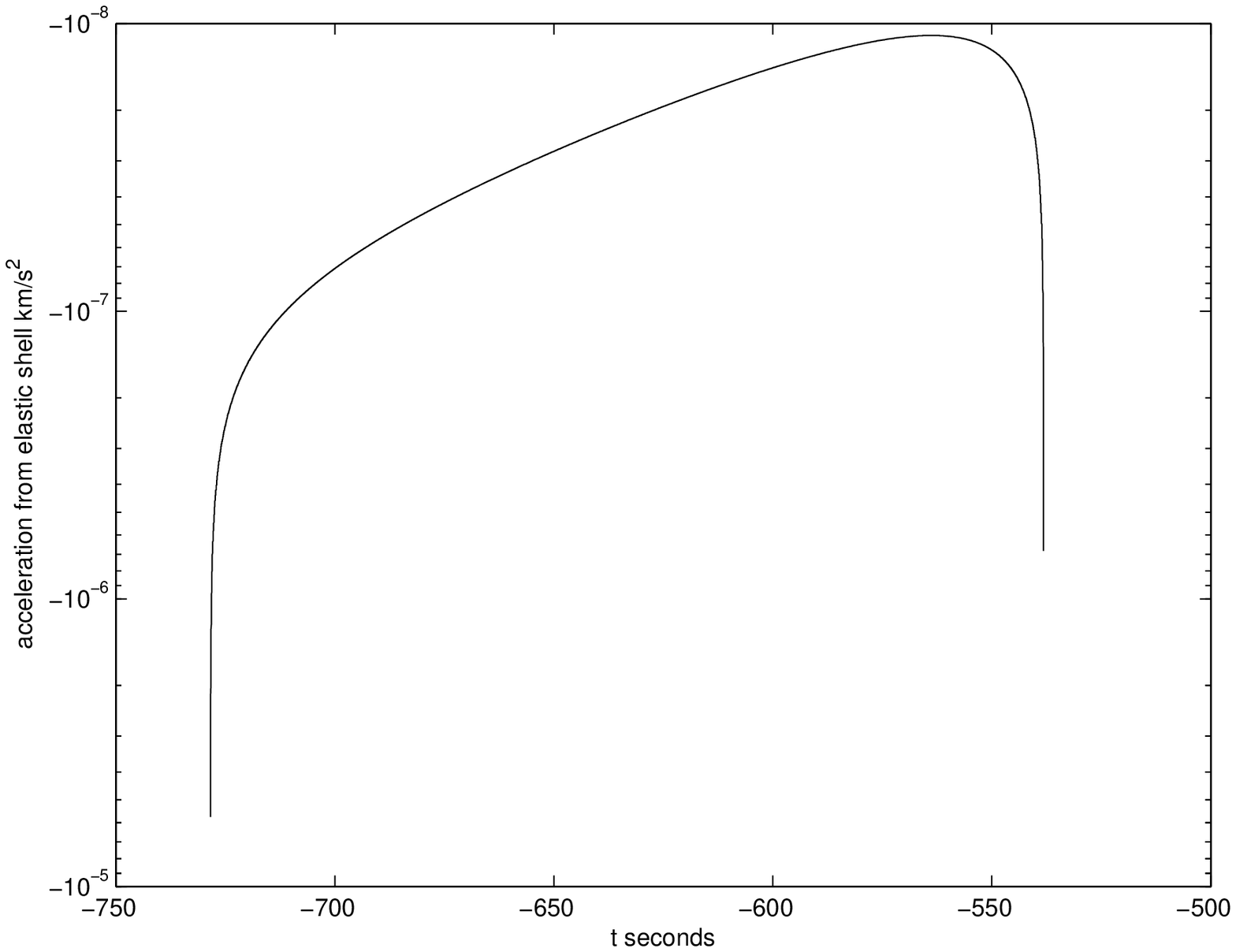}
\caption{GLL-II downtrack acceleration from elastic shell ${\rm km}/{\rm s}^2$ }
\end{figure}

\vfill\eject

\begin{figure}
\includegraphics[angle=-90]{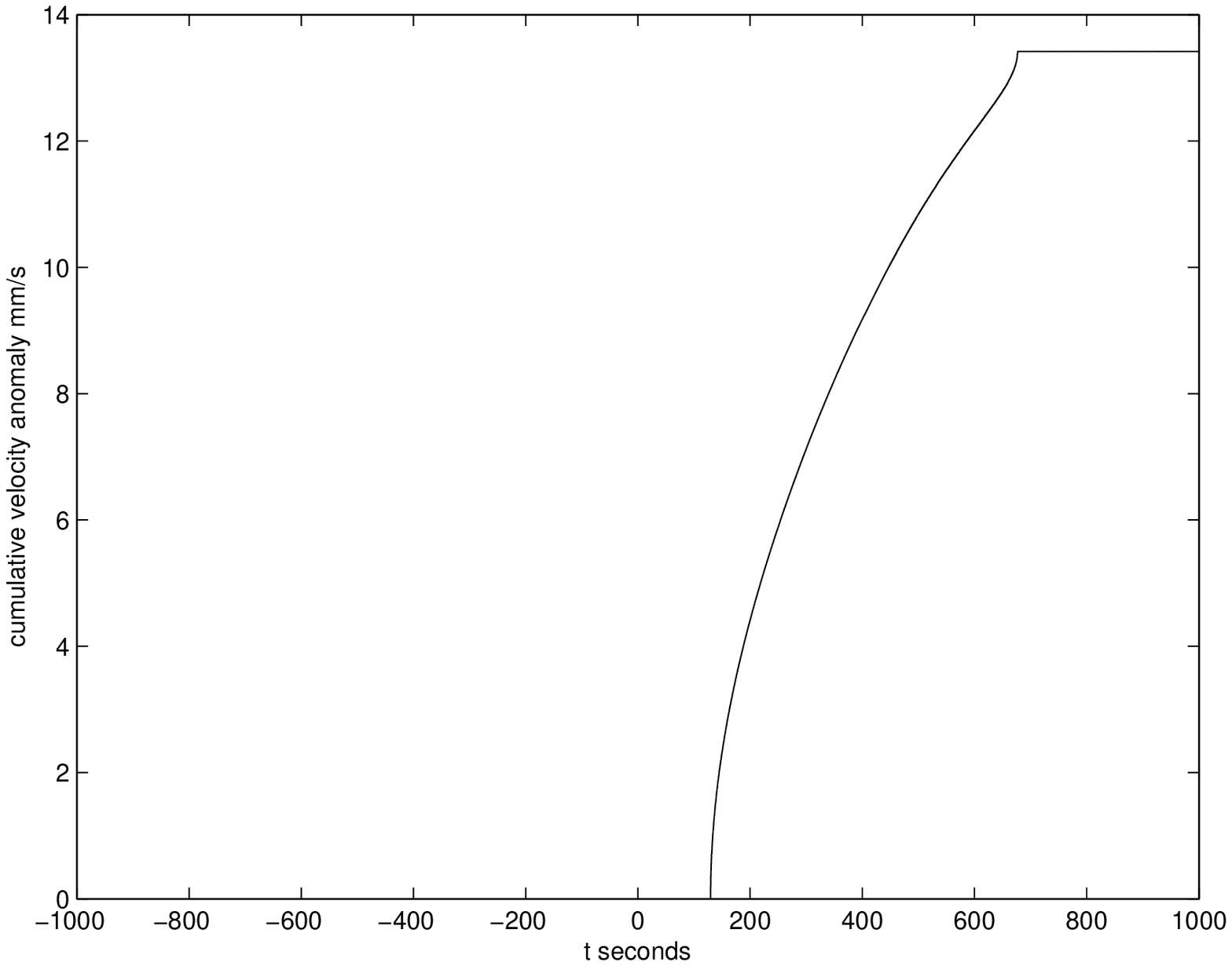}
\caption{NEAR cumulative asymptotic velocity anomaly $\delta V_{\infty}$  in ${\rm mm}/{\rm s}$}
\end{figure}

\vfill\eject

\begin{figure}
\includegraphics[angle=-90]{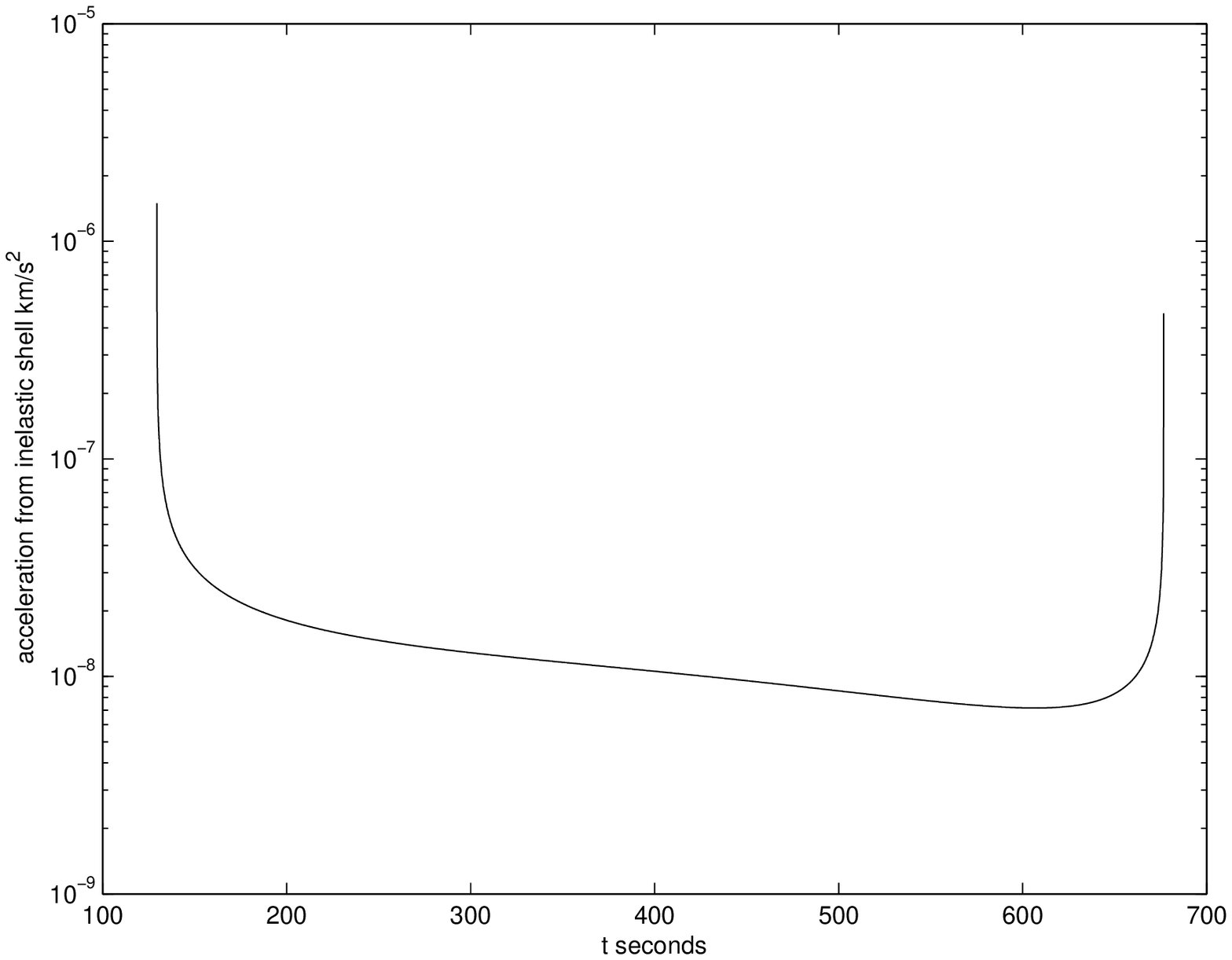}
\caption{NEAR downtrack acceleration from inelastic shell ${\rm km}/{\rm s}^2$ }
\end{figure}

\vfill\eject

\begin{figure}
\includegraphics[angle=-90]{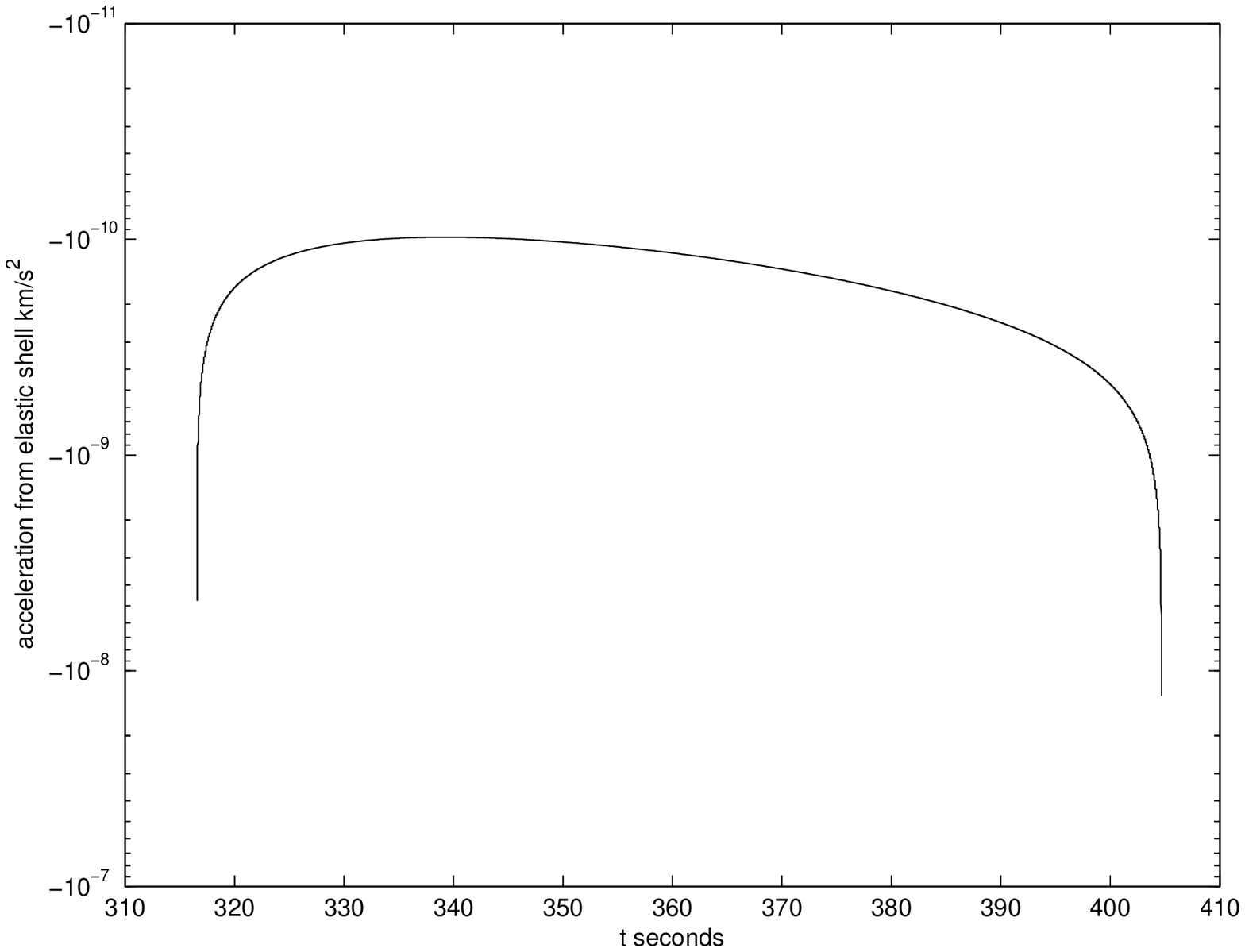}
\caption{NEAR downtrack acceleration from elastic shell ${\rm km}/{\rm s}^2$ }
\end{figure}

\vfill\eject

\begin{figure}
\includegraphics[angle=-90]{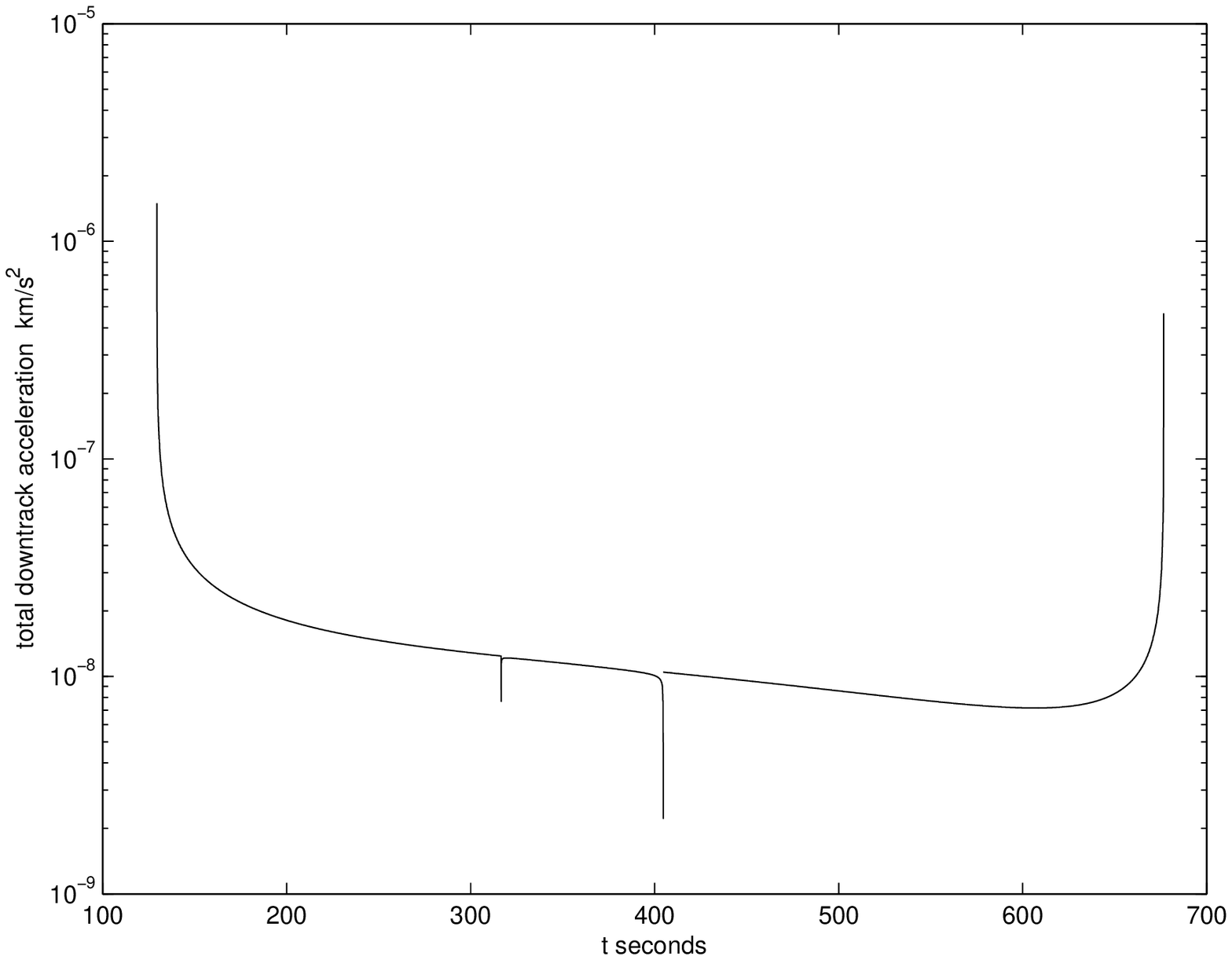}
\caption{NEAR total downtrack acceleration ${\rm km}/{\rm s}^2$ }
\end{figure}

\vfill\eject

\begin{figure}
\includegraphics[angle=-90]{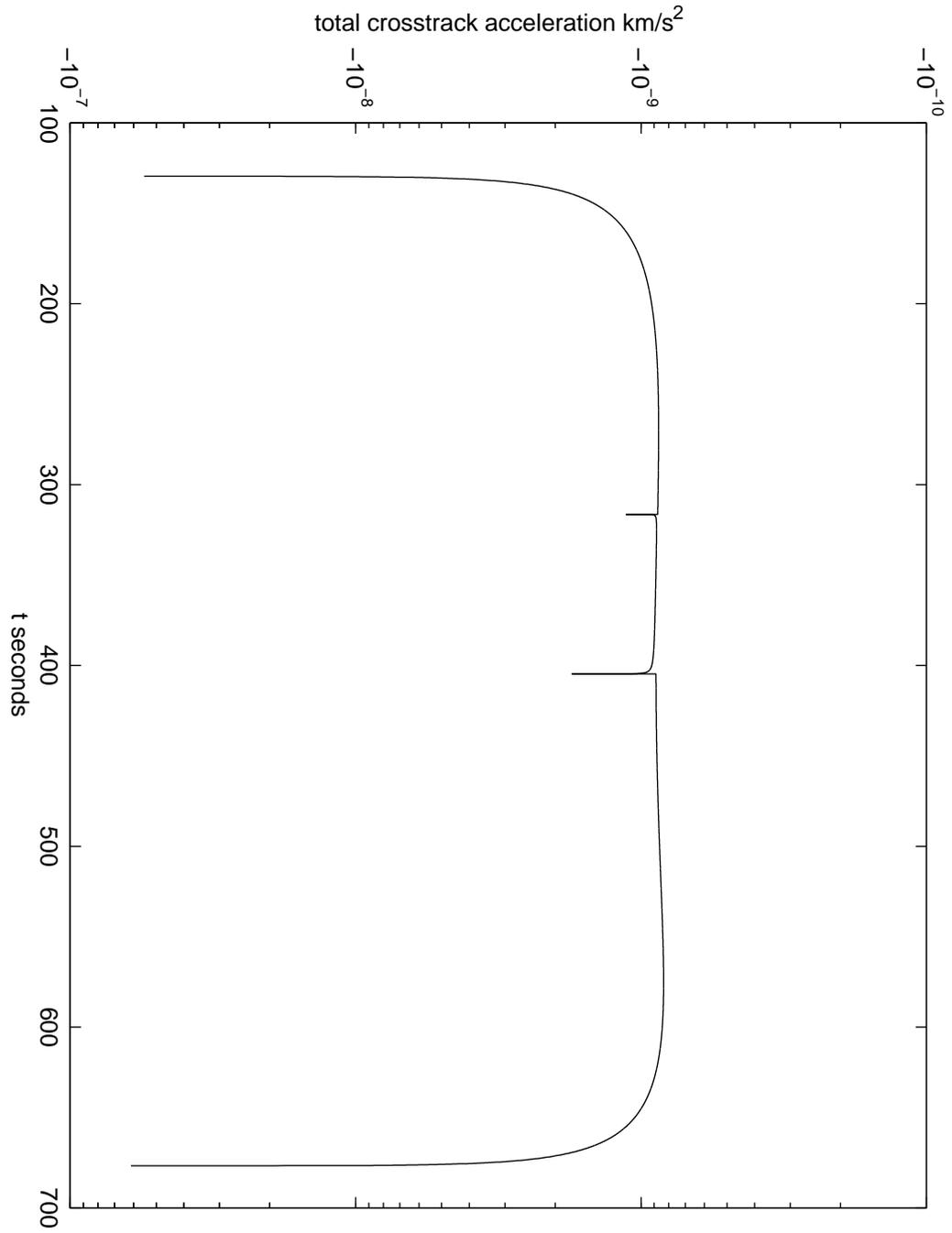}
\caption{NEAR total crosstrack acceleration ${\rm km}/{\rm s}^2$ }
\end{figure}

\vfill\eject

\begin{figure}
\includegraphics[angle=-90]{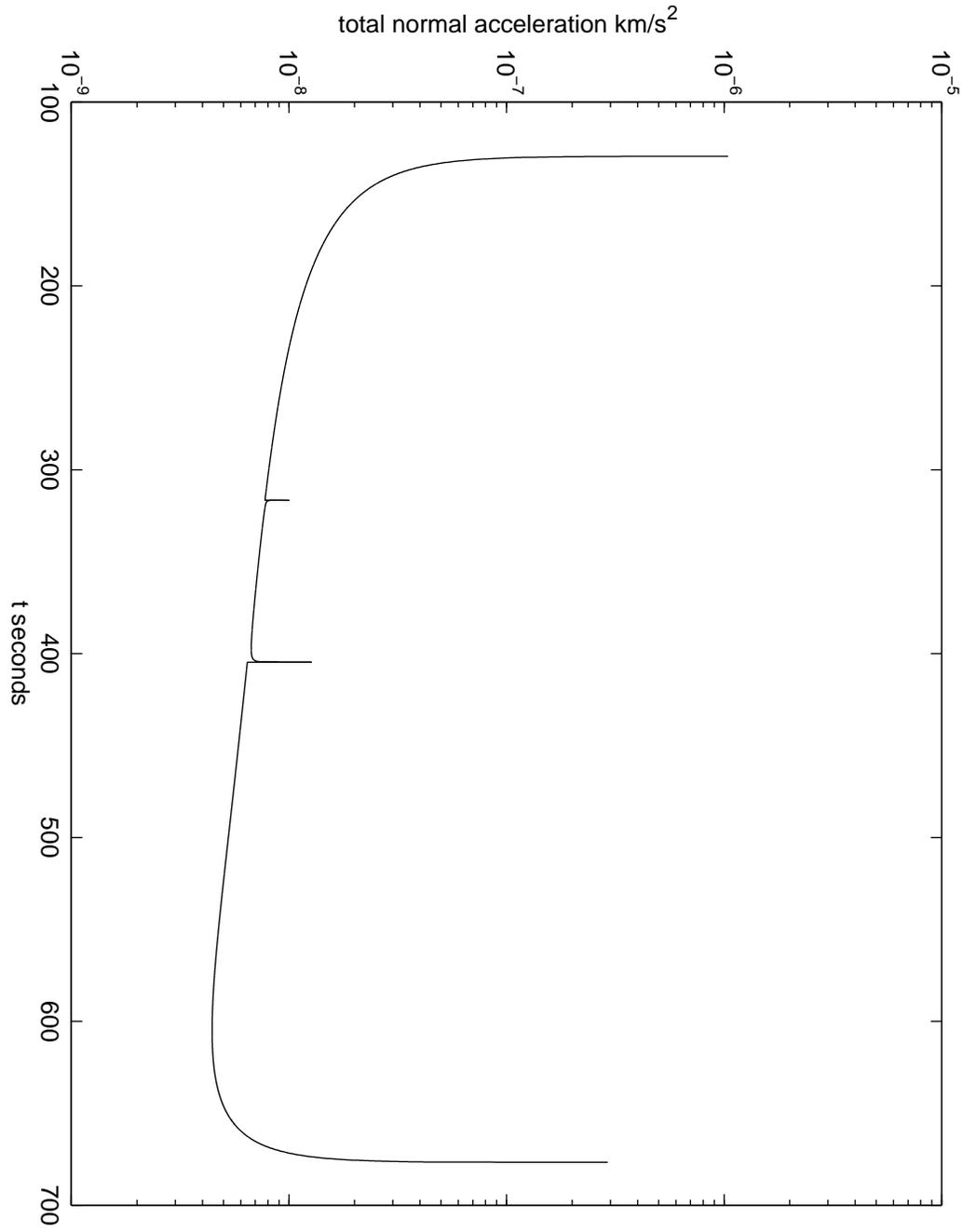}
\caption{NEAR total acceleration normal to orbit plane ${\rm km}/{\rm s}^2$ }
\end{figure}

\vfill\eject

\begin{figure}
\includegraphics[angle=-90]{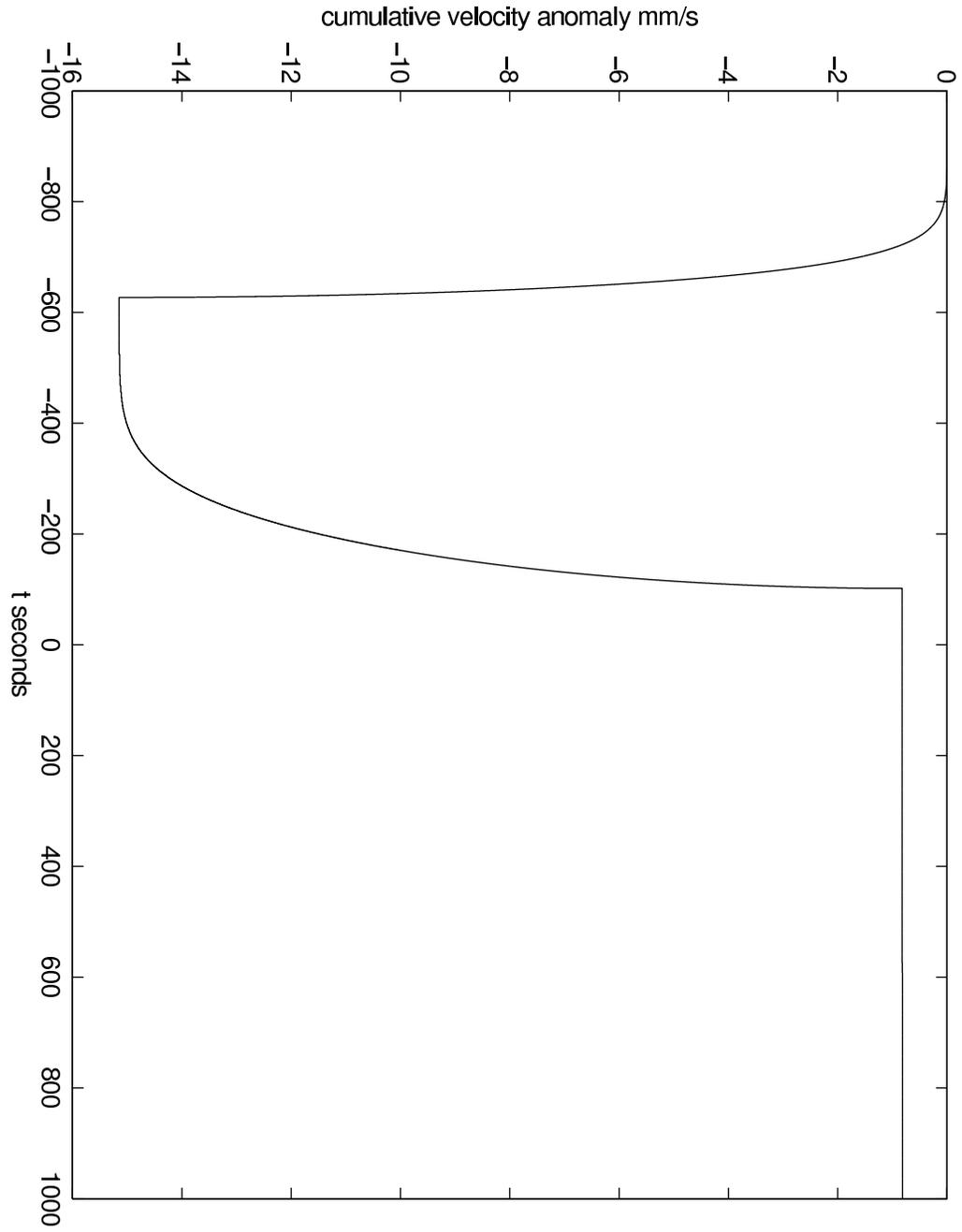}
\caption{Cassini cumulative asymptotic velocity anomaly $\delta V_{\infty}$  in ${\rm mm}/{\rm s}$}
\end{figure}

\vfill\eject

\begin{figure}
\includegraphics[angle=-90]{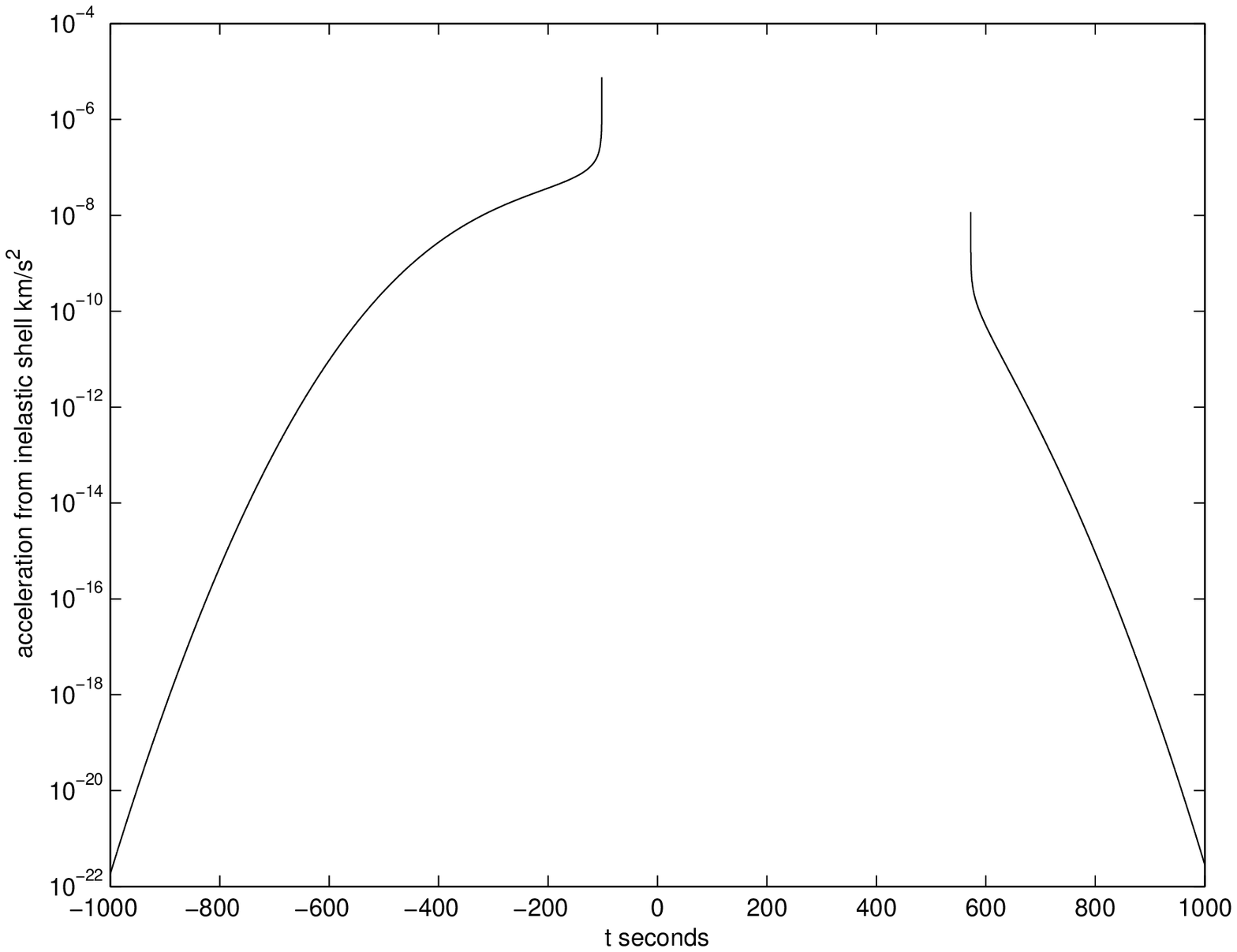}
\caption{Cassini downtrack acceleration from inelastic shell ${\rm km}/{\rm s}^2$ }
\end{figure}

\vfill\eject

\begin{figure}
\includegraphics[angle=-90]{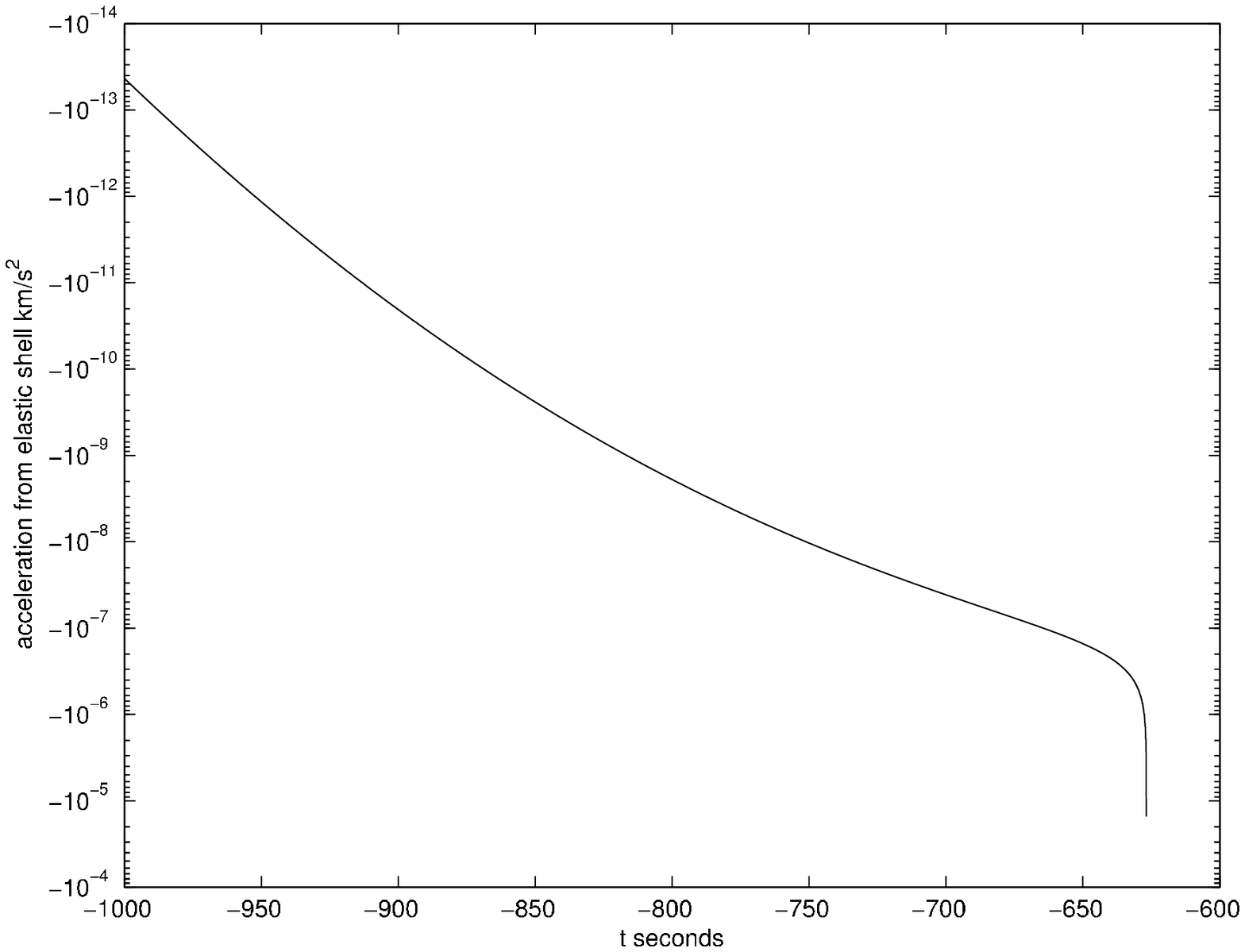} 
\caption{Cassini downtrack acceleration from elastic shell ${\rm km}/{\rm s}^2$ }
\end{figure}

\vfill\eject

\begin{figure}
\includegraphics[angle=-90]{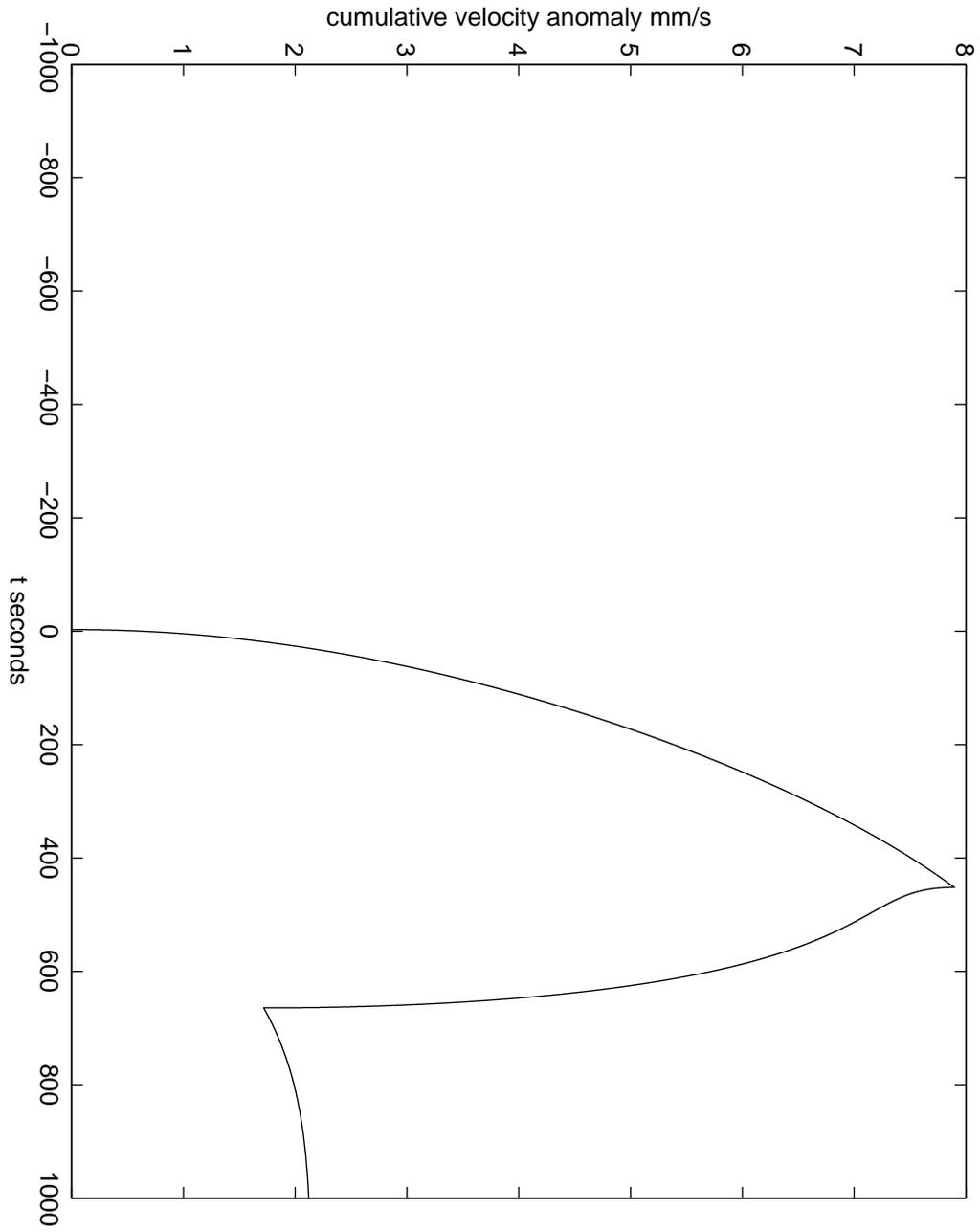}
\caption{Rosetta cumulative asymptotic velocity anomaly $\delta V_{\infty}$  in ${\rm mm}/{\rm s}$}
\end{figure}

\vfill\eject

\begin{figure}
\includegraphics[angle=-90]{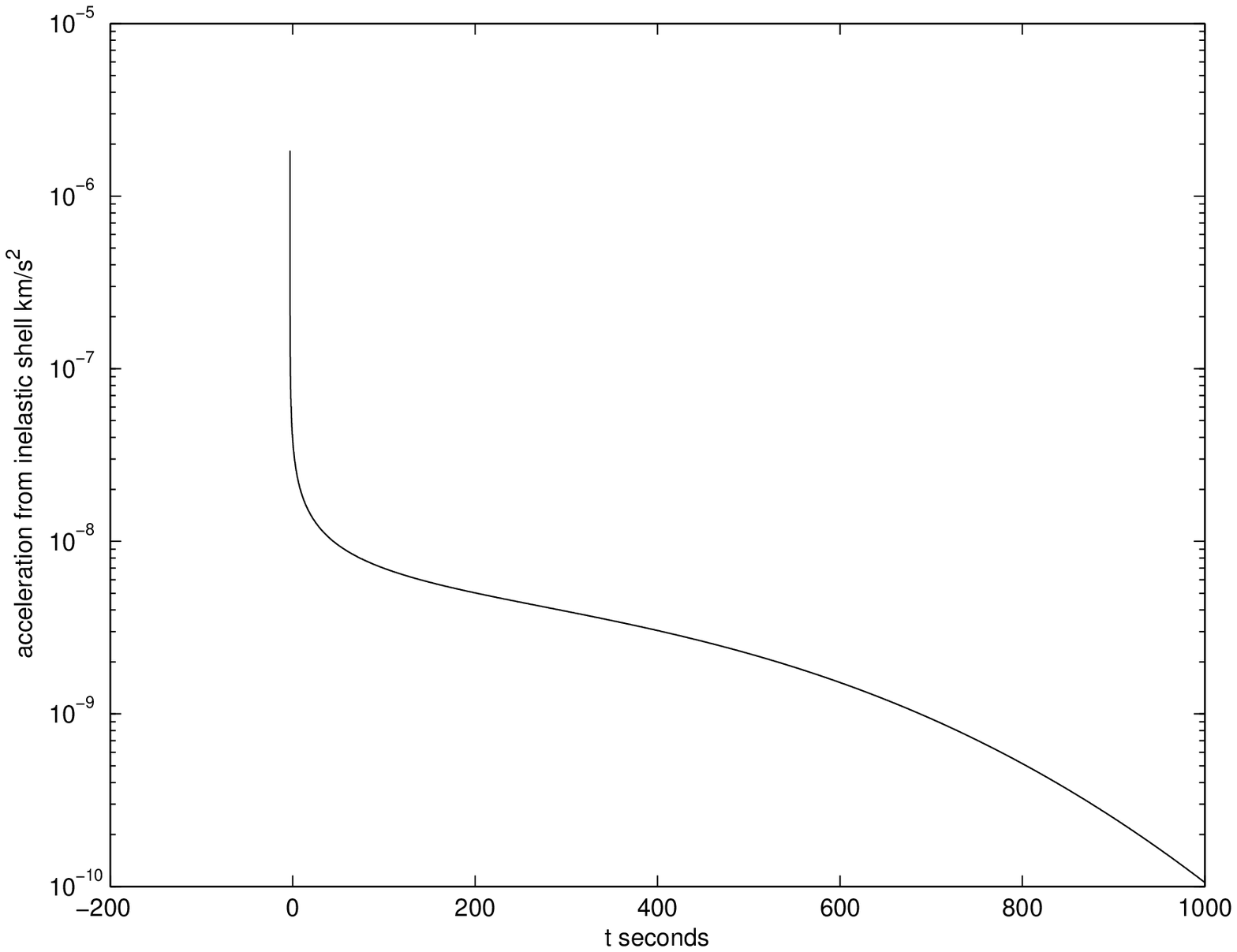}
\caption{Rosetta downtrack acceleration from inelastic shell ${\rm km}/{\rm s}^2$ }
\end{figure}

\vfill\eject

\begin{figure}
\includegraphics[angle=-90]{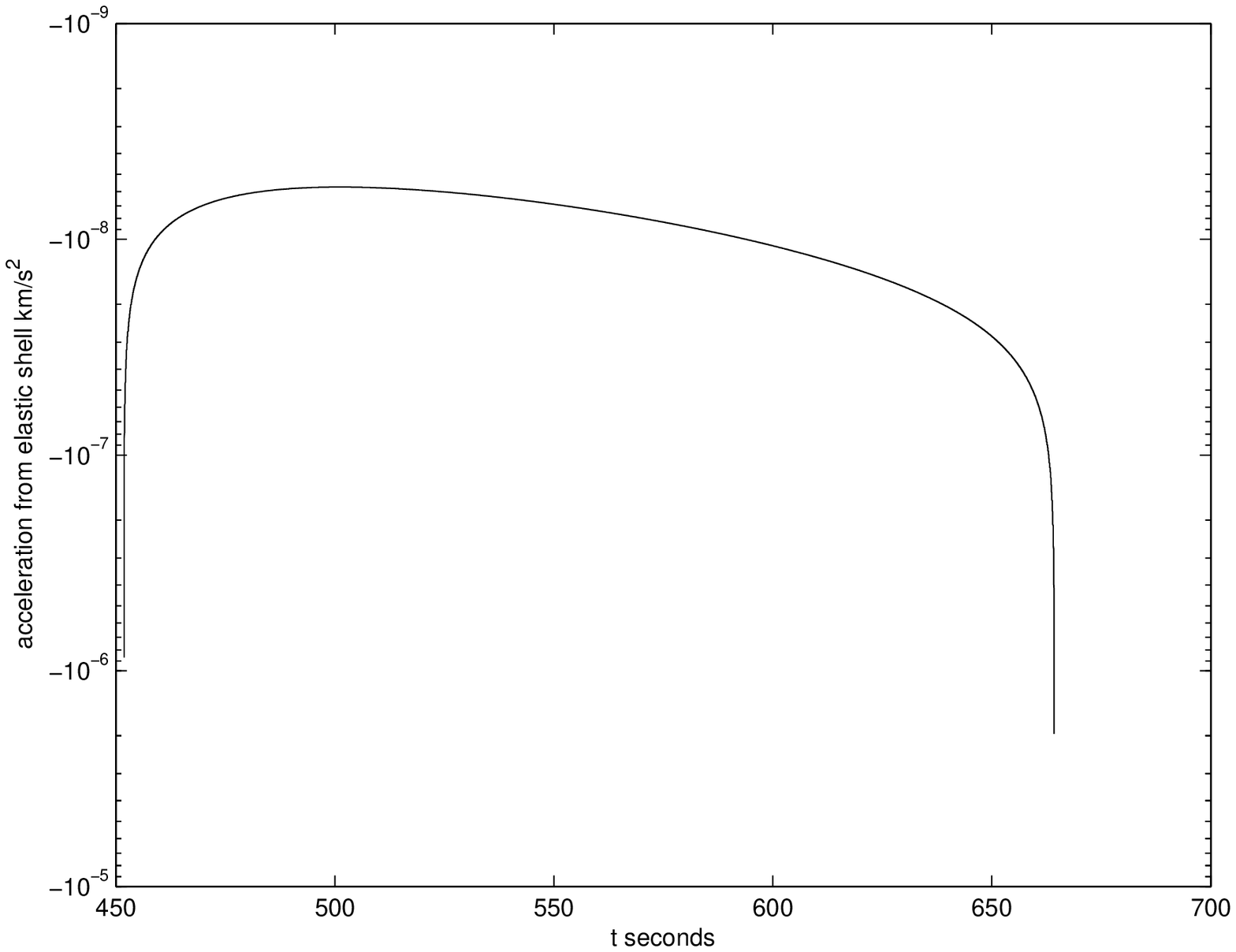}
\caption{Rosetta downtrack acceleration from elastic shell ${\rm km}/{\rm s}^2$ }
\end{figure}

\vfill\eject

\begin{figure}
\includegraphics[angle=-90]{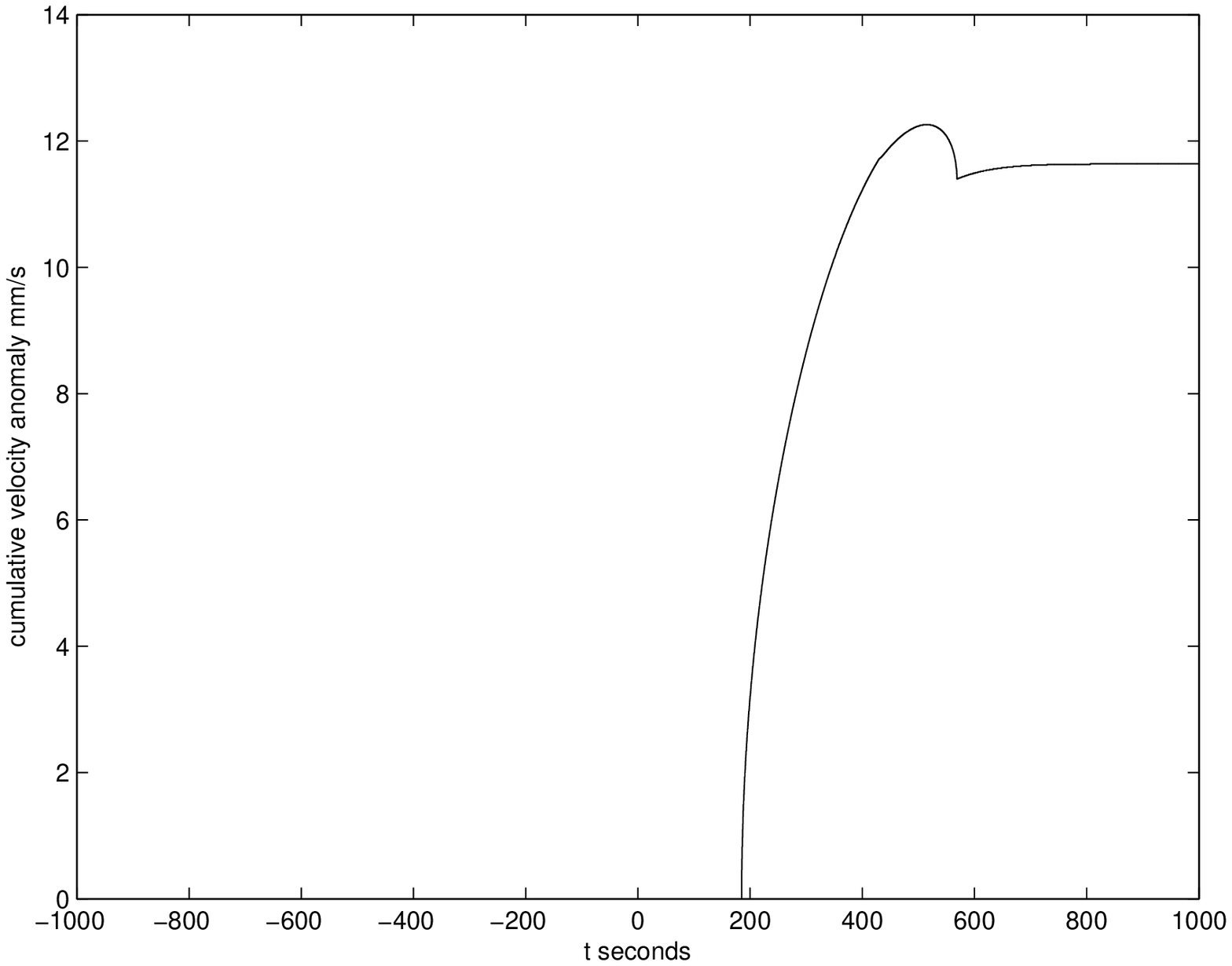}
\caption{Juno cumulative asymptotic velocity anomaly $\delta V_{\infty}$ in ${\rm mm}/{\rm s}$}
\end{figure}

\vfill\eject

\begin{figure}
\includegraphics[angle=-90]{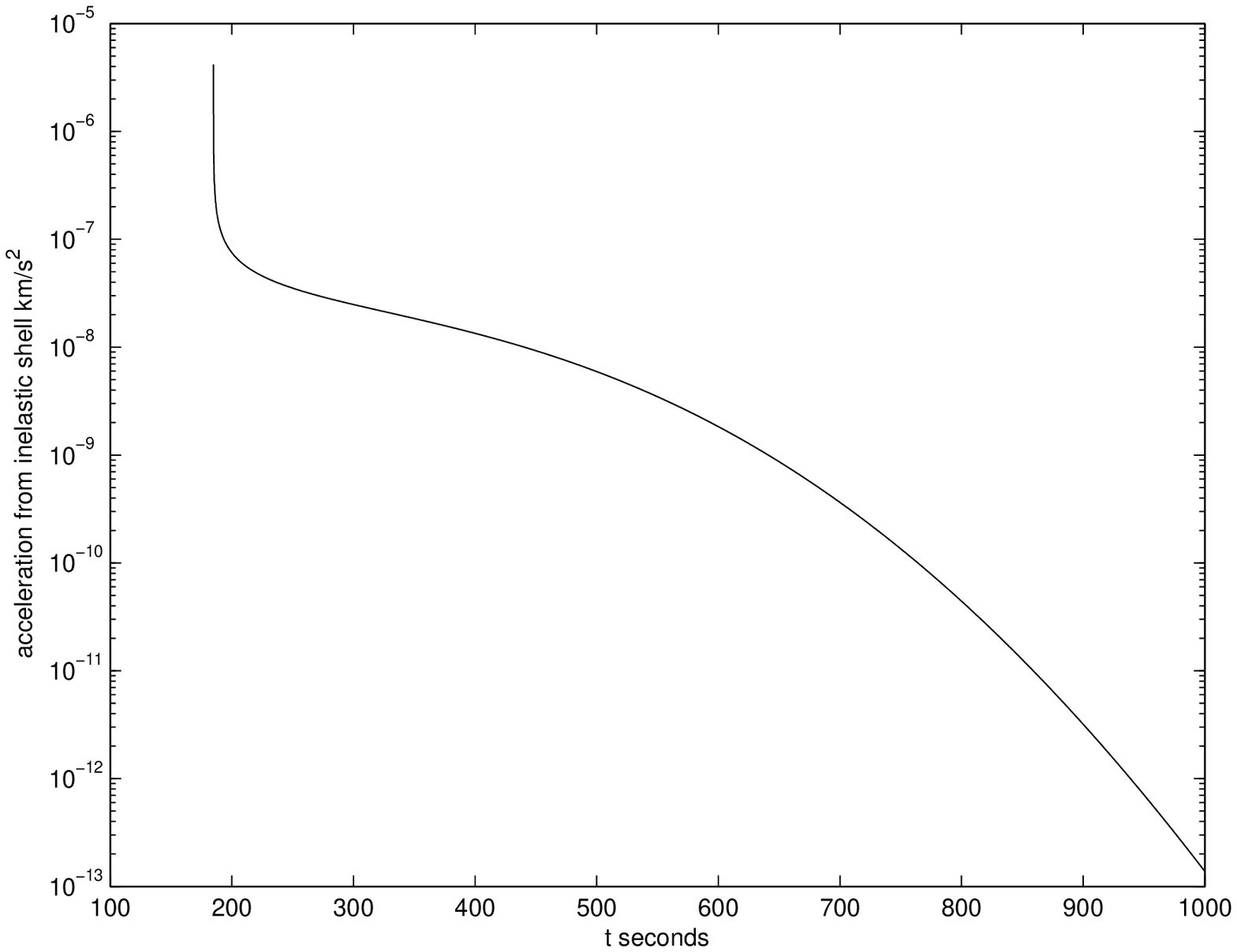}
\caption{Juno downtrack acceleration from inelastic shell ${\rm km}/{\rm s}^2$ }
\end{figure}

\vfill\eject

\begin{figure}
\includegraphics[angle=-90]{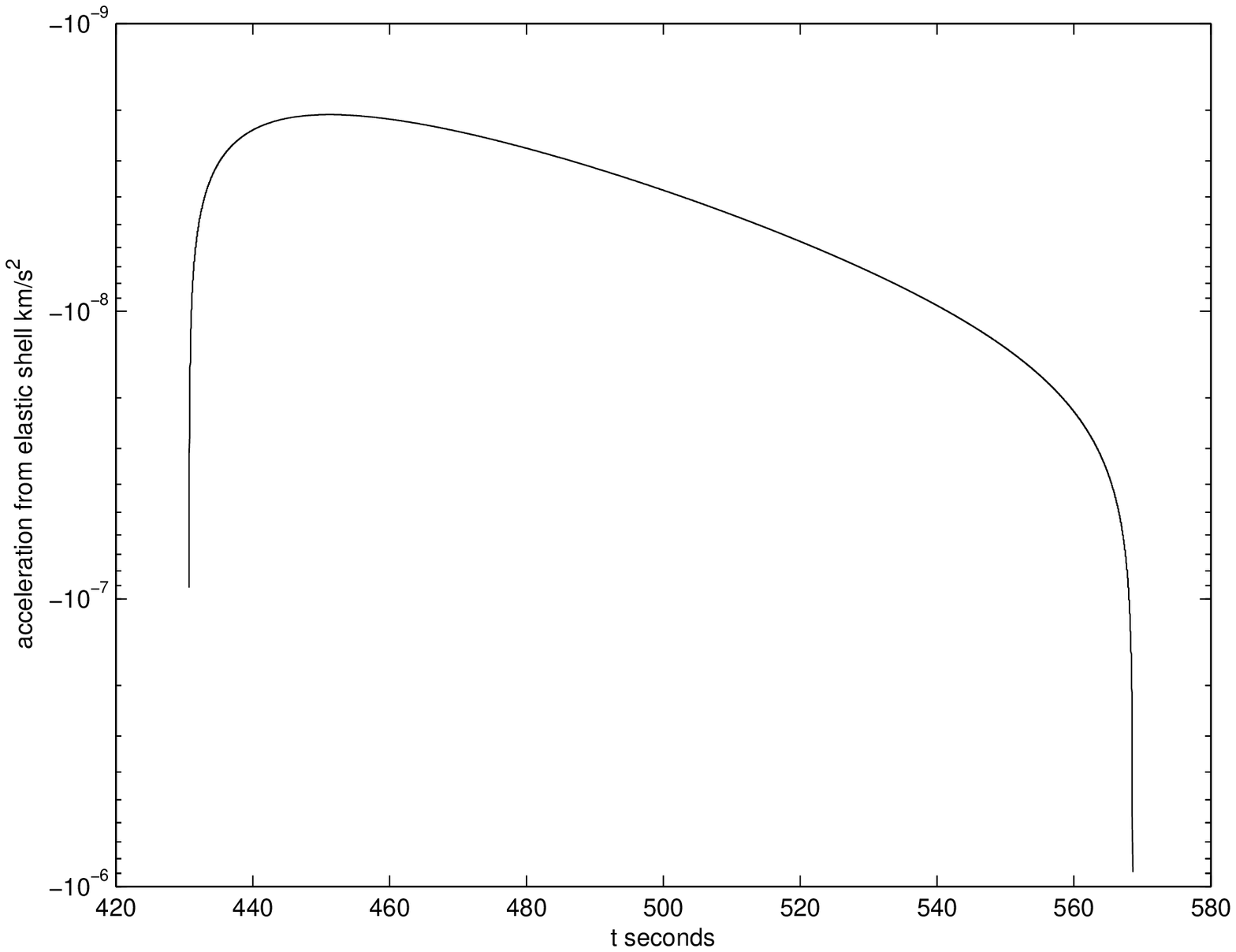}
\caption{Juno downtrack acceleration from elastic shell ${\rm km}/{\rm s}^2$ }
\end{figure}

\vfill\eject

\vfill\break


\begin{thebibliography}    {99}
\bibitem{adler1} S. L. Adler,  Phys. Rev. D {\bf 79}, 023505 (2009).

\bibitem{adler2}  S. L. Adler, ``Modeling the flyby anomalies with dark matter scattering'', in H. Fritzsch and K. K. Phua, eds.,
Proceedings of the Conference in Honour of Murray Gell-Mann's 80th Birthday, World Scientific (2011), and in a more expanded version,
arXiv:0908.2412.

\bibitem{adler3}  S. L. Adler, J. Phys. A:Math Theor. {\bf 41} (2008) 412002.

\bibitem{adler4} S. L. Adler, ``Spacecraft calorimetry as a test of the dark matter scattering model for flyby anomalies'',
arXiv:0910.1564. Version 1 of this posting was  submitted as a white paper to the NAS decadal review on biological and physical sciences in space.

\bibitem{anderson}  J. D. Anderson, J. K. Campbell, J. E. Ekelund, J. Ellis,
and J. F. Jordan, Phys. Rev. Lett. {\bf 100}, 091102 (2008).

\bibitem{andersoncampbell}  J. D. Anderson and J. K. Campbell, private email communications (2011).


\end{thebibliography}
\end{document}